\documentstyle[epsfig,12pt]{article}

\newlength{\dinwidth}                                                  
\newlength{\dinmargin}                                                 
\def\lapproxeq{\lower .7ex\hbox{$\;\stackrel{\textstyle                
<}{\sim}\;$}}
\def\gapproxeq{\lower .7ex\hbox{$\;\stackrel{\textstyle                
>}{\sim}\;$}}
\def\beq{\begin{equation}}
\def\eeq{\end{equation}}
\def\bea{\begin{eqnarray}}
\def\eea{\end{eqnarray}}
\def\rhob{\bar \rho}
\def\etab{\bar \eta}
\def\eps{\epsilon}
\def\epsb{\bar \epsilon}

\newcounter{omi}
\newcounter{add}

\newcommand{\ord}[1]{{\mathcal{O}}\left( #1 \right)}
\newcommand{\vev}[1]{\left\langle #1\right\rangle}

\newcommand{\srr}{s^{\prime }{}_{23}^{Y}}
\newcommand{\srd}{s^{\prime }{}_{13}^{D}}
\newcommand{\srdd}{s^{\prime }{}_{23}^{D}}
\def\circa#1{\,\raise.3ex\hbox{$#1$\kern-.75em\lower1ex\hbox{$\sim$}}\,}
\newlength{\myem}
\newcounter{mysubequation}[equation]

\begin{document}

\title{\begin{flushright}\small                                                      
OUTP-01-23-P  \\                                                       
RAL-TR-2001-007 \\
FERMILAB-Pub-01/028-T \\
{\tt hep-ph/0104088}\\
\end{flushright}{\Large {\bf Precision test of a Fermion mass texture}}}
\author{
R.\ G.\ Roberts$^{(a)}$,
A.\ Romanino$^{(b)}$, 
G.\ G.\ Ross$^{(c)}$ and
L.Velasco-Sevilla$^{(c)}$ \bigskip 
\and $^{(a)}${\small Rutherford Appleton Laboratory, Chilton, Didcot,
  Oxon, OX11 0QX, UK} 
\and $^{(b)}${\small Fermi National Accelerator Laboratory P.O. Box 500,
Batavia, IL 60510, USA} 
\and $^{(c)}${\small Department of Physics, University of Oxford,
  Oxford, OX1 3NP, UK}}
\maketitle

\begin{abstract}
  Texture zeros in the quark Yukawa matrices generally lead to precise
  and simple expressions for CKM matrix elements in terms of ratios of
  quark masses. Using the new data on $b-$decays we test a
  particularly promising texture zero solution and show that it is at
  best approximate. We analyse the approximate texture zero
  structure and show it is consistent with experiment. 
  We investigate the implications for the
  CKM unitarity triangle, measurements at $BaBar$ and $BELLE$ as
  well as for the theories which invoke family symmetries.
\end{abstract}

\section{\noindent {\protect\large {\bf Introduction}}}

The structure of quark and lepton mass matrices provides us with a rare
insight into the physics beyond the Standard Model which may directly probe
the underlying theory at the gauge unification or Planck scale. While the
quark mass matrices and the CKM matrix, $V^{CKM},$ are intimately related,
measurement of the eigenvalues of the mass matrices and the matrix elements
of $V^{CKM}$ is not sufficient to determine the structure of the full mass
matrix and of the matrix of Yukawa couplings giving rise to them. Given this
under-determination, the phenomenological approach most often used is to
make some assumption about this structure and explore the experimental
consequences for the $V_{ij}^{CKM}$. A particularly promising starting point
assumes that there are anomalously small entries in the up and down quark
Yukawa matrices - ``texture zeros''\footnote{%
Strictly texture zeros can only apply at a single mass scale (the GUT or
string scale?) and will be filled in by Renormalisation Group running.
However, in general, such effects are very small and the texture zeros
persist to a good approximation at all scales.}. \ These lead to relations
for the $V_{ij}^{CKM}$ in terms of ratios of quark masses which do not
involve any unknown couplings and hence can be precisely tested. Various
texture zeros have been studied. For the case of symmetric mass matrices a
systematic analyses determining which combinations of textures involving 4,
5 or 6 zeros for the $U$, $D$ matrices are compatible with data was carried
out in \cite{rrr}. The main reason for looking for such texture zero
solutions is that they may shed light on physics beyond the Standard Model,
for example the presence of a new family symmetry relating different
generations.

The new generation of b-factory experiments has led to more precise
measurements of the CKM\ matrix elements that, together with the
progress in understanding hadronic
uncertainties~\cite{Ciuchini:2000de} and light quark masses, allows us
to test texture zero structures to a greater precision than has
hitherto been possible. In this paper we will study the most promising
structure based on simultaneous zeros in the $U$ and $D$ mass matrices
at the (1,1) and (1,3) positions.  In the additional hypothesis of
equal magnitude of the (1,2) and (2,1) entries and of sufficiently
small (3,2) entry, this structure gives three precise relations
between $V_{ij}^{CKM}$ and ratios of quark masses, leaving only one
CKM element undetermined. We find that the new experimental and
theoretical information suggests that (at least) one of the hypothesis
leading to those precise texture zero relations needs to be relaxed.
The simplest possibility is that while the (1,3) element is small it
is {\it non-} zero so that the texture zero is only approximate.  As a
result one of the relations, the texture zero prediction for
$|V_{ub}/V_{cb}|$, is modified, as is suggested by the new precise
data; the other two relations are less affected.

 In the context of an
underlying family symmetry this result is to be expected for the
family symmetry usually requires texture zeros to be only approximate
and, in some cases, actually predicts the order at which the
approximate zero should be filled in.  For example, a very simple
Abelian family symmetry predicts the (1,3) element should be nonzero
at a level consistent with the new data while requiring the (1,1)
element should be much smaller, preserving the remaining two texture
zero predictions. 

Another possibility~\cite{Romanino:2000dz} is that the (3,2) entry in the
down quark mass matrix is not as small as the (2,3) entry
(barring cancellations, the latter must be smaller than the (3,3)
entry by a factor $\ord{|V_{cb}|}$). Such an asymmetry can also be
easily achieved in the context of Abelian or non-Abelian family
symmetries and offers an intriguing connection with neutrino physics.
A sizeable (3,2) entry in the down quark mass matrix is in fact a
generic prediction of a class of unified models of quark and lepton
masses and mixings. In such models, a large leptonic mixing angle
accounting for the atmospheric neutrino anomaly originates from a
sizeable (2,3) entry in the charged lepton mass matrix. The GUT
symmetry then forces the (3,2) element in the down quark matrix to be
of the same order of magnitude~\cite{Albright:1998vf}.  
However we note that the original symmetric structure can also lead,
very naturally, to a large neutrino mixing angle~\cite{tasi}.  
Finally, the
third situation leading to a correction to the texture zero relations
is that the entries (1,2) and (2,1) in the up quark mass matrix are
not equal in magnitude. This can happen in non unified abelian models
due to different order one coefficients. We will not investigate this
possibility in this paper because it destroys one of the successful
texture zero predictions (see the discussion in Section \ref{tests}
below).  
From this one sees that detailed tests of
the texture zero relations will help to identify the underlying family
symmetry.

Our analysis does not take into account possible new physics
contributions to the processes constraining the CKM parameters.
Such contributions might affect the experimental determination of
$|V_{td}/V_{ts}|$ through modifications to $B$ mixing and the $CP$-violating
part of $K$ mixing (or both) and consequently affect the corresponding texture 
zero relation. However we do not expect $|V_{ub}/V_{cb}|$ to be similarly
affected where the main uncertainty instead lies in the hadronic modelling of
charmless semileptonic $B$ decays.
Our analysis addresses the phenomenological problem of the $|V_{ub}/V_{cb}|$ 
prediction of texture zero structures and its remedies and,
as a consequence, most of our analysis would not be affected by new
physics effects. Very recent measurements of $\sin 2\beta$ 
from $BaBar$ and $BELLE$~\cite{BB:2001} 
do raise the possibility of beyond Standard Model contributions to 
$CP$-violation and therefore  
the quantitative fits performed
in Sections \ref{tests} and \ref{pert} might be affected by
supersymmetric contributions to $K$ and $B$ mixing. We briefly discuss how
low values of $\sin 2\beta$ affect our analysis.

The paper is structured as follows. In Section \ref{tests} we present
the experimental tests of the texture zero predictions following from
zeros in the $U$ and $D$ mass matrices at the (1,1) and (1,3)
positions. In Section \ref{pert} we discuss the implications for the
mass matrices following from the need to modify one of the texture
zero relations. We develop a perturbative expansion which allows us to
identify the possible corrections to the texture zero predictions.
In Section \ref {family} we consider the implications of such
structure for an underlying family symmetry and finally in Section 5 we
present conclusions.

\section{{\protect\large {\bf Tests of texture zero predictions\label{tests}}%
    }}

In what follows we assume that the off-diagonal entries are small
relative to their on-diagonal partners so that one may develop a
perturbative expansion for the CKM matrix elements \cite{bj,hall,
  ggr,fritzsch}. This is a reasonable starting point because it
immediately leads to small mixing angles consistent with observation.
We further assume here that there are texture zeros in the (1,1) and
(1,3) elements, that the (1,2) and (2,1) elements have equal magnitude
and that the texture is approximately symmetric ((3,2)$\sim$(2,3)).
These assumptions lead to the texture zero relations
\cite{rrr,hall,barbieri,fritzsch,gsto}
\begin{equation}
\left| \frac{V_{ub}}{V_{cb}}\right| =\sqrt{\frac{m_{u}}{m_{c}}}\quad \quad
\quad \quad \left| \frac{V_{td}}{V_{ts}}\right| =\sqrt{\frac{m_{d}}{m_{s}}}
\label{eq:vub1}
\end{equation}
\begin{equation}
|V_{us}|=\lambda =\left| \sqrt{
 \frac{m_{d}}{m_{s}}}-e^{i\phi}\sqrt{\frac{m_{u}}{m_{c}}} \right|
 \label{eq:vus1}
\end{equation}
We will prove that, quite generally, $\phi $ is approximately the Standard
Model CP violating phase (for more restricted cases see \cite
{hall1,fritzsch1}).

Barbieri et al (BHR) \cite{barbieri} emphasized that these relations lead to
a very tight determination of the CKM unitarity triangle. In terms of the
re-scaled Wolfenstein parameters, $\bar{\rho}=c\rho $, $\bar{\eta}=c\eta $, $%
c=(1-\lambda ^{2}/2),$ the ratios $|V_{ub}|/|V_{cb}|$ and $|V_{td}|/|V_{ts}|$
are 
\begin{equation}
\frac{|V_{ub}|}{|V_{cb}|}=\frac{\lambda }{c}\sqrt{\bar{\rho}^{2}+\bar{\eta}%
^{2}}\;\;\;\;\;\;\;\;\;\frac{|V_{td}|}{|V_{ts}|}=\frac{\lambda }{c}\sqrt{(1-%
\bar{\rho})^{2}+\bar{\eta}^{2}}  \label{eq:vubcbtdts}
\end{equation}

In the context of the Standard Model, the measurable quantities which give
information on $\bar{\rho}$ and $\bar{\eta}$ are \newline
\indent(a) the ratio $|V_{ub}|/|V_{cb}|$ obtained from semi-leptonic decays
of $B$ mesons, \newline
\indent(b) $\Delta m_{B_{d}}$ and $\Delta m_{B_{s}}$ which are the mass
differences in the $B_{d}^{0}-\bar{B^{0}}_{d}$, $B_{s}^{0}-\bar{B^{0}}_{s}$
systems, \newline
\indent(c) $|\epsilon _{K}|$ the parameter related to CP violation in the $K$%
, $\bar{K}$ system, and \newline
\indent(d) $\sin 2\beta $ ($\beta $ is one of the angles in the unitarity
triangle) obtained from CP asymmetries in various $B$ decays.

\begin{table}[t]
\label{tablesminF}
\par
\begin{center}
\begin{tabular}{|r|l|c|}
\hline
\multicolumn{3}{|c|}{Fixed Parameters}\\ \hline
\hline
Parameter & Value  & Reference \\ 
$G_{F}$ & $1.16639\times 10^{-5}{\rm GeV}^{-2}$ & \cite{pdgb:dat} \\ 
$M_{W}$ & $(80.42\pm 0.06){\rm GeV}$ & \cite{pdgb:dat} \\ 
$f_{K}$ & $(0.161\pm 0.0015){\rm GeV}$ & \cite{pdgb:dat} \\ 
$m_{K}$ & $(0.497672\pm 0.000031){\rm GeV}$ & \cite{pdgb:dat} \\ 
$\Delta m_{K}$ & $(3.491\pm 0.009)\times 10^{15}{\rm GeV}$ &
\cite{pdgb:dat}
\\ 
$|\epsilon _{K}|$ & $(2.271\pm 0.017)\times 10^{-3}$ & \cite{pdgb:dat} \\ 
$\eta _{2}^{\star }$ & $(0.574\pm 0.004)$ & \cite{caravparrodst:99} \\ 
$m_{B_{d}}$ & $(5.2792\pm 0.0018){\rm GeV}$ & \cite{pdgb:dat} \\ 
$\eta _{B}$ & $0.55\pm 0.01$ & \cite{caravparrodst:99} \\ 
$m_{B_{s}}$ & $(5.3693\pm 0.0020){\rm GeV}$ & \cite{pdgb:dat} \\ \hline
\end{tabular}
\end{center}
\caption{\small{Fixed Parameters. }}
\label{fixed}
\end{table}

\subsection{Standard Model (SM) fit}
\label{SMfit}

In comparing the texture zeros with experiment we proceed in two stages. We
first use the latest data to find $\lambda $, $\bar{\rho}$ and $\bar{\eta}$
and then compare the result with eq(\ref{eq:vubcbtdts}). Our procedure is to
construct a two-dimensional probability density for $\bar{\rho}$ and $\bar{%
\eta}$ \cite{Paganini:1997cu} from the constraints of the above
measurements. Entering in the fits are several parameters which have been
well measured and which we choose not to vary. These are given in Table \ref
{fixed}. At present we have only an upper limit for $\Delta m_{B_{s}}$ and
we employ the so-called `amplitude method' to include this information into
the fit\cite{hgmoserros:97}.

Our fit assumes the Standard Model (SM) relations between the
experimental measurables and the CKM matrix elements. In a
supersymmetric extension of the Standard Model we expect that there
will be corrections to these measurables; $|\epsilon _{K}|$ is
particularly sensitive to such corrections. To allow for this
possibility we carry out a separate fit in which the data on
$|\epsilon _{K}|$, together with that on $\sin 2\beta $ for which
there are rather different measurements, is dropped. The mass
differences $\Delta m_{B_d}$, $\Delta m_{B_s}$ and $\sin 2\beta$ might
also be affected by supersymmetry, especially if the structure of
squark mass matrices is determined by the flavour symmetry accounting
for quark masses and mixings. We do not consider this possibility
here.

The formulas used in the standard model fit are 
\begin{equation}
\Delta m_{B_{d}}=C_{\Delta m_{B_{d}}}A^{2}\lambda ^{6}[(1-\bar{\rho})^{2}+%
\bar{\eta}^{2}]m_{B_{d}}f_{B_{d}}^{2}B_{B_{d}}\eta _{B_{d}}S(x_{t}^{\star }),
\label{eq:DeltamBdWolfpar}
\end{equation}
where $C_{\Delta m_{B_d}}=\frac{G_{F}^{2}M_{W}^{2}}{6\pi ^{2}}$. Here $%
S(x_{t}^{\star })$ is the standard Inami-Lim function \cite{inamilim} and $%
f_{B}^{2}B_{B}$ is the product of the $B$ meson decay constant and the $B$
parameter analogous to $B_{K}$ in the $K$ system. 
\begin{equation}
\Delta m_{B_{s}}=\Delta m_{B_{d}}\frac{m_{B_{s}}}{m_{B_{d}}}\xi ^{2}\frac{%
c^{2}}{\lambda ^{2}}\frac{1}{(1-\bar{\rho})^{2}+\bar{\eta}^{2}},
\label{DeltamBsWolfp}
\end{equation}
where 
\begin{equation}
\xi =\frac{f_{B_{s}}\sqrt{B_{B_{s}}}}{f_{B_{d}}\sqrt{B_{B_{d}}}}.
\end{equation}

For the $\epsilon _{K}$ parameter we have
\begin{eqnarray} 
|\epsilon _{K}| &=&C_{\epsilon }B_{K}A^{2}\lambda^{6} 
\bar{\eta}\left[-\eta_{1}^{\star } x_{c}^{\star }+A^{2}\lambda^{4}
\left( 1-\bar{\rho}-\left(  
\bar{\rho}^{2}+\bar{\eta}^{2}-\rhob\right) \lambda^{2}\right) 
\eta_{2}^{\star }S(x_{t}^{\star }) \right. \nonumber \\
&+&\left. \eta _{3}^{\star }S(x_{c}^{\star },x_{t}^{\star })\right],
\label{epsKwolfpar} 
\end{eqnarray}
where $C_{\epsilon }=\frac{G_{F}^{2}f_{K}^{2}m_{K}m_{W}^{2}}{6\sqrt{2}\pi
^{2}\Delta m_{K}}$. The short distance QCD corrections are contained in the
coefficients $\eta _{i}^{\star }$, which have been computed at the Next to
Leading-logarithmic Order (NLO) in \cite{Buras:1990fn}. The NLO calculation
requires the use of the one-loop relation between the pole mass and the
running mass $m_{i}^{pole}=m_{i}^{\star }\left( 1+\frac{\alpha _{s}(m^{\star
})}{\pi }\frac{4}{3}\right) $ in the $\overline{{\rm MS}}$. The
coefficients $\eta _{i}^{\star }$ have been evaluated at $\Lambda _{%
\overline{MS}}^{{NLO}}$=$371MeV$. All the starred quantities in Table \ref
{varied} are given in terms of these quantities, for example $x_{i}^{\star
}=[m^{\star }]^{2}/M_{W}^{2}$.

Finally the angles of the unitarity triangle are given by 
\begin{eqnarray}
\sin 2\beta &=&\frac{2\bar{\eta}(1-\bar{\rho})}{\bar{\eta}^{2}+(1-\bar{%
\rho})^{2}}  \nonumber \\
\sin 2\alpha &=&\frac{2\bar{\eta}\left( \bar{\eta}^{2}+\bar{\rho}(\bar{%
\rho}-1)\right) }{\left( \bar{\eta}^{2}+(1-\bar{\rho})^{2}\right) \left( 
\bar{\eta}^{2}+\bar{\rho}^{2}\right) }  \nonumber \\
\sin 2\gamma &=&\frac{2\bar{\rho}\bar{\eta}}{\bar{\rho}^{2}+\bar{\eta}^{2}%
}  \label{eq:sin2beta}
\end{eqnarray}

Using these expressions we carried out fits to the parameters listed in
Table \ref{varied} keeping the well determined parameters listed in Table 
\ref{fixed} fixed.

\begin{table}[ht]
\label{tablesminV}
\begin{center}
\begin{tabular}{|r|l|c|c|} \hline
\multicolumn{4}{|c|}{Fitted Parameters}\\ \hline \hline
Parameter & Value & Gaussian-Flat errors. &Referen.\\
$A$ & $0.834\pm 0.036$ & & * \\
$\lambda$ & $0.2196\pm 0.0023$ & & *\\
$|V_{ub}|^{{\rm CLEO}}$ &$32.5\times 10^{-4}$ &$(\pm 2.9 \pm 5.5)\times
10^{-4}$ & \cite{cleoVub:99} \\
$|V_{ub}|^{{\rm LEP}}$ &$41.3\times 10^{-4}$ & $(\pm 6.3 \pm 3.1)\times
10^{-4}  $ & \cite{LEPwo:2000}\\
$|V_{cb}|$ &$(41.0\pm 1.6)\times 10^{-3}$& & \cite{Ciuchini:2000de}\\
$B_K$ & $0.87\pm 0.143$ &$0.06\pm 0.13$ &\cite{caravparrodst:99}\\
$m_c^{\star}$ & $(1.3\pm 0.1) {\rm GeV}$ & &*\\
$m_t^{\star}$ & $(167\pm 5){\rm GeV}$ & & *\\
$\eta_1^{\star}$ & $1.38\pm 0.53 $ & &\cite{caravparrodst:99}\\
$\eta_3^{\star}$ & $0.47\pm 0.04$ & &\cite{caravparrodst:99}\\
$\Delta m_{B_d}$ & $(0.487\pm 0.014){\rm ps}^{-1}$ &
&\cite{lepbosonline:00} \\
$f_{B_d}\sqrt{B_{B_d}}$ & $(0.230\pm 0.032){\rm GeV}$ & $\pm 0.025$-$ \pm
0.020$ & \cite{Ciuchini:2000de}\\
$\xi$& $(1.14\pm 0.064)$ &$0.04\pm0.05$\ &\cite{Ciuchini:2000de}\\
$\Delta m_{B_s}>$ & $15{\rm ps}^{-1}$ at 95\% C.L.&
&\cite{lepbosonline:00} \\
$\sin 2\beta$ & $0.41\pm 0.17$ &   &\cite{BB:2001} \\
$\sin 2\beta$ & $0.47\pm 0.16$ &   &\cite{Wu:2001qu} \\
\hline
\end{tabular}
\end{center}
\caption{\small {Fitted Parameters.The parameters marked with  * have been
computed here with the new data from~\protect\cite{pdgb:dat}.}}
\label{varied}
\end{table}

\begin{figure}[ht]
\vspace*{-0.75cm}
\begin{center}
\mbox{\epsfig{file=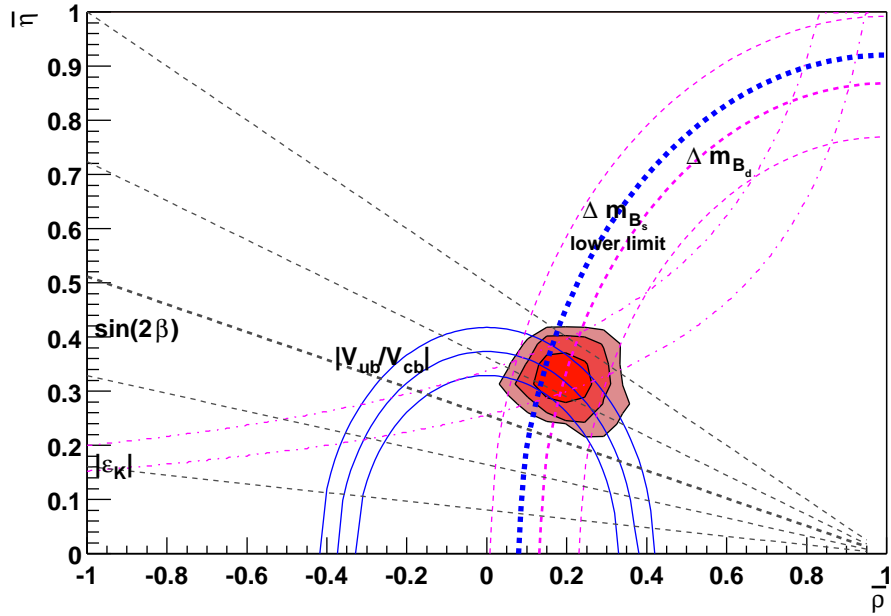,height=9cm}}
\end{center}
\vspace*{-0.75cm}
\caption{\small{The SM fit of Section 2.1 to $|V_{ub}/V_{cb}|$, 
$\Delta m_{B_s}$ (lower limit), $\Delta m_{B_d}$, $|\epsilon_K|$ 
and the recent result for $\sin 2\beta$. The lines indicate
the region of 1$\sigma$ and, in the case of $\sin 2\beta$,
also the 2$\sigma$ region, demonstrating that the new value is
still consistent with the rest of SM constraints within 2$\sigma$. The
CL are at 99\%, 95\% and 68\%.}}
\label{fig:fclsmo1}
\end{figure}

The results in the $\bar{\rho}-\bar{\eta}$ plane of the fit are shown in
Fig.\ref{fig:fclsmo1}.
The confidence limits shown correspond to 68\%, 95\% and 99\%. In
the experimental fits (which assume the SM) the results of
$|V_{ub}/V_{cb}|$ for both $CLEO$ and $LEP$ collaborations (see Table
\ref{varied}) were included.  Although the result from the $CLEO$
collaboration is lower than the result from $LEP$, they are consistent
within one $\sigma $. The combination of both, assuming a Gaussian
distribution for the experimental errors and a flat distribution for
the theoretical errors, gives a value of $0.087\pm 0.010$, which is
consistent with the PDG value of $0.090\pm 0.025$
\cite{pdgb:dat}.

As can be seen from Fig \ref{fig:fclsmo1} the constraints on 
$\bar{\rho}$ and $\bar{\eta}$ would be considerably strengthened by a
measurement of $\Delta m_{B_{s}}$, which presently has only a lower limit,
and by an improvement in the precision on $\sin 2\beta$.
We can see from Fig.\ref{fig:fclsmo1} this agrees within $2\sigma$ with the
the new value for the parameter $\sin 2\beta$=$0.47\pm0.16$ 
(combining $BaBar$ and $Belle$ results with those from $CDF$ and $ALEPH$
~\cite{Wu:2001qu}). However there is
a deviation at the $1 \sigma$ level which may be a hint for physics beyond
the SM.

\subsection{Comparison with the texture zero predictions.}
\label{comparison}

\begin{figure}[ht] 
\vspace*{-0.75cm}
\begin{center}
\mbox{\epsfig{file=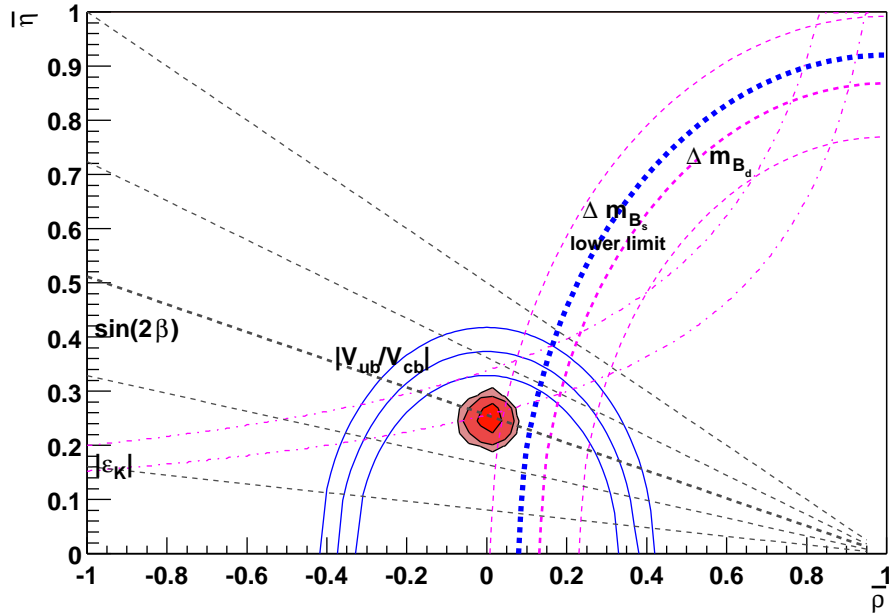,height=9cm}}
\end{center}
\vspace*{-0.75cm}
\caption{\small{The fit for the relations 
$|V_{td}|/|V_{ts}|=~\protect\sqrt{m_d/m_s}$
and $|V_{ub}|/V_{cb}|=~\protect\sqrt{m_u/m_c}$. 
The experimental constraints + SM interpretation indicate that
these relations are disfavoured.}}
\label{fig:fclst1}
\end{figure}

We are now able to compare the experimental results with the texture
zero predictions of eq(\ref{eq:vub1}). In Fig.\ref{fig:fclst1} we show 
the region in the $\bar{\rho}-\bar{\eta}$ plane allowed by these relations
together with the various constraints following from the processes
(a)-(d). In this fit we have taken symmetric forms for the $U$ and $D$ 
Yukawa matrices with texture zeros in the (1,1), (1,3) and (3,1) positions.
Comparison with Fig \ref{fig:fclsmo1} show that the 
predictions are hard to reconcile with
the data, being consistent only at greater than the 99\% CL (for the case $%
|\epsilon _{K}|$ and $\sin 2\beta $ are not included). 
Fig.\ref{fig:fclst1}
shows that one of the reasons for the poor agreement is the measurement of $%
\left| \frac{V_{ub}}{V_{cb}}\right| .$ Given the fact the CLEO\ and LEP\
measurements differ considerably it is of interest to consider whether the
discrepancy disappears if we use only the value for the $CLEO$
collaboration. In fact we find that this only marginally changes things. 
One may see that the significant improvement in the
experimental measurements, particularly of $\left| \frac{V_{ub}}{V_{cb}}%
\right| $ and the improved lower limit on $\Delta m_{B_{s}}$, strongly
disfavour this promising texture zero scheme.  As remarked before,
this conclusion holds in extensions of the SM too, since new physics
effects might affect the mass difference $\Delta m_{B_{s}}$, but
cannot alleviate the disagreement with $V_{ub}/V_{cb}$.

Given this discrepancy, do we have to abandon the texture zero
solution completely? In fact we do not, as we now show. The
  problematic relation $|V_{ub}/V_{cb}|=\sqrt{m_u/m_c}$ (and, to a
  lesser extent, the other two texture zero predictions) depends on
  three assumptions.
  \begin{itemize}
  \item Texture zeros: the matrix elements $Y_{13}$, $Y_{31}$,
    $Y_{11}$ are negligibly small both in the up ($Y=U$) and down
    ($Y=D$) sector. Actually, as mentioned above, the underlying
    theory generating the texture zero is unlikely to guarantee a
    particular mass matrix element is absolutely zero.  Moreover an
    exact zero can only apply at a single mass scale for radiative
    effects necessarily generate contributions to all the elements of
    the quark mass matrices.
  \item Small higher order corrections in the perturbative
    diagonalization of the up and down quark mass matrices. 
    The expected correction is of order 7\% or less
    if $|D_{32}|\sim |V_{cb}||D_{33}|$ and becomes important for
    larger values of $|D_{32}|$. Larger values of $|D_{32}|$ can arise
    in unified models in which a large leptonic mixing originates from
    the charged lepton sector. They also arise in models with a
    texture zero in the (2,2) position and no cancellation between
    down and up quark contributions to $V_{cb}$. In this case one has
    in fact $|D_{23}/D_{33}| \sim |V_{cb}|$. As a consequence, an
    asymmetry $|D_{32}/D_{33}|> |D_{23}/D_{33}|$ is required in order
    to account for the value of $m_s/m_b = |D_{23}/D_{33} \cdot
    D_{32}/D_{33}|$.
  \item $|U_{12}|=|U_{21}|$. This condition is usually met in unified
    models. In the case of SU(5) models for example, $U_{12}$ and
    $U_{21}$ are both generated by operators that can be written in
    the form $\vev{\phi}T_1 T_2 H$, where $T_{1,2}$ are the tenplets
    of the first and second family containing the up quarks, $H$ is
    the up Higgs fiveplet and $\phi$ represents a (normalized) set of
    fields whose vev is SM invariant. The relation $|U_{12}|=|U_{21}|$
    follows unless $\vev{\phi}$ breaks SU(5). In the latter case,
    SU(5) Clebsh coefficients will differentiate $|U_{12}|$ and
    $|U_{21}|$, but often in a too violent way, leading to an even
    worse disagreement with the $V_{ub}/V_{cb}$ prediction.  Moreover,
    the analogous operator in the down quark and charged lepton sector
    would spoil the successful relation $m_d m_s/m_b^2 \sim m_e
    m_\mu/m_\tau^2$. The condition $|U_{12}|=|U_{21}|$ is
    automatically met in some non-Abelian
    models~\cite{Barbieri:1996uv,Barbieri:1997tu}. In any case we regard
    the phenomenological success of the relation for $|V_{us}|$ given by
    eq(\ref{eq:vus1}) as a result that should be preserved.

\end{itemize}
In what follows we focus on the possibility that either the first
or the second assumption is not fulfilled. In the first case we still assume
a symmetric structure for $U$ and $D$ and since in the second case $D$ is 
manifestly asymmetric, we refer to the these two scenarios  as the 
{\it symmetric} and {\it asymmetric} texture cases.  

In particular, in
Section~\ref{sec:u1} we study in detail textures with small but
non negligible (1,3) element. As we discuss there, the order at
which this element arises is a characteristic prediction of a family
symmetry so determination of this element is a discriminator between
various candidate family symmetries. Moreover we show that, even if we
drop the constraint of an exact texture zero in the (1,3)
position, two of the three texture zero predictions remain and are in
good agreement with the data.  Finally we prove that the
identification of the phase $\phi $ in the expression for $V_{us}$,
eq(\ref{eq:vus1}), is true in leading order even after allowing for a
matrix element in the (1,3) position.  In~\ref{sec:u2} we 
consider the complementary possibility of an asymmetric texture.

\section{Non-zero $s_{13}$ : Perturbative Analysis\label{pert}}

In this Section we use the notation of Hall and Rasin \cite{hall}. We
start with the Yukawa matrices {\bf Y} ({\bf Y}={\bf U} or {\bf D})
with the assumption that the entries in the Yukawa matrices have a
hierarchical structure, with $Y_{33}$ being the largest. For the
purposes of explaining the important aspects of the analysis it is
useful to first take $Y_{ij}$ to be real and later consider how the
analysis is modified by CP violating phases. The matrices {\bf Y} can
be diagonalized by three successive rotations in the (2,3), (1,3) and
(1,2) sectors (denoted by $s_{23},s_{13}$ and $s_{12}$ ):

\begin{eqnarray}
\lefteqn{\left( \begin{array}{ccc} \widetilde{\widetilde{Y}}_{11} & 0 & 0 \\
0 & \widetilde{\widetilde{Y}}_{22} & 0 \\ 0 & 0 & \widetilde{\widetilde{%
Y}}_{33} \end{array} \right) =\left( \begin{array}{ccc} 1 & -s{}_{12}^{Y} & 0
\\ s{}_{12}^{Y} & 1 & 0 \\ 0 & 0 & 1 \end{array} \right) \left( \begin{array%
}{ccc} 1 & 0 & -s{}_{13}^{Y} \\ 0 & 1 & 0 \\ s{}_{13}^{Y} & 0 & 1 \end{array}
\right) \left( \begin{array}{ccc} 1 & 0 & 0 \\ 0 & 1 & -s{}_{23}^{Y} \\ 0 &
s{}_{23}^{Y} & 1 \end{array} \right) \times }  \nonumber \\
&&\times \left( 
\begin{array}{ccc}
Y_{11} & Y_{12} & Y_{13} \\ 
Y_{21} & Y_{22} & Y_{23} \\ 
Y_{31} & Y_{32} & Y_{33}
\end{array}
\right) \left( 
\begin{array}{ccc}
1 & 0 & 0 \\ 
0 & 1 & s^{\prime }{}_{23}^{Y} \\ 
0 & -s^{\prime }{}_{23}^{Y} & 1
\end{array}
\right) \left( 
\begin{array}{ccc}
1 & 0 & s^{\prime }{}_{13}^{Y} \\ 
0 & 1 & 0 \\ 
-s^{\prime }{}_{13}^{Y} & 0 & 1
\end{array}
\right) \left( 
\begin{array}{ccc}
1 & s^{\prime }{}_{12}^{Y} & 0 \\ 
-s^{\prime }{}_{12}^{Y} & 1 & 0 \\ 
0 & 0 & 1
\end{array}
\right) .  \label{rot}
\end{eqnarray}
In terms of these angles the CKM\ matrix is given by 
\begin{equation}
{\bf V}=\left( 
\begin{array}{ccc}
1 & s_{12}+s_{13}^{U}s_{23} & s_{13}-s_{12}^{U}s_{23} \\ 
-s_{12}-s_{13}^{D}s_{23} & 1 & s_{23}+s_{12}^{U}s_{13} \\ 
-s_{13}+s_{12}^{D}s_{23} & -s_{23}-s_{12}^{D}s_{13} & 1
\end{array}
\right) ,  \label{vhr}
\end{equation}
where $s_{23}=s_{23}^{D}-s_{23}^{U}$, $s_{13}=s_{13}^{D}-s_{13}^{U}$ and $%
s_{12}=s_{12}^{D}-s_{12}^{U}$.

It is straightforward now to see the origin of the texture zero
relations eq( \ref{eq:vub1}). From eq(\ref{vhr}) we see it is
sufficient to have \cite{hall}:

$\bullet $ ${\frac{|V_{ub}|}{|V_{cb}|}}=|s_{12}^{U}|$ and ${\frac{|V_{td}|}{%
|V_{ts}|}}=|s_{12}^{D}|$ which is obtained by: 
\begin{equation}
|s_{13}|<<|s_{12}^{U}s_{23}|\,\,{\rm and}\,\,|s_{13}|<<|s_{12}^{D}s_{23}|.
\label{tz13}
\end{equation}
This condition will define how small the (1,3) element must be in the
mass matrices to obtain the (1,3) ``texture zero'' prediction.  Notice
that since the ``13'' rotations are performed after the ``23''
rotations, what determines the size of the rotation $s_{13}$ is the
``effective'' element $\tilde Y_{13}$ in eq.~(\ref{tilde13}).  As
eq.~(\ref{tilde13}) shows, $\tilde Y_{13}$ depends not only on
$Y_{13}$ but also on the size of the $Y_{21}$ element, which is
rotated in the (3,1) position by the right-handed quark rotation
$\srr$. The size of the $Y_{21}$ element can be related to the light
quark masses, so that the conditions~(\ref{tz13}) also defines how
small the right-handed rotation $\srr$ must be in order to obtain the
texture zero prediction.

In addition we require

$\bullet$ $|s^{U}_{12}| = \sqrt{\frac{ m_u }{m_c }}$ and $|s^{D}_{12}| = 
\sqrt{\frac{ m_d }{m_s }}$ which is obtained by:

\begin{equation}
|{\widetilde{Y}}_{11}|<<|{\frac{{{\widetilde{Y}}_{12}{\widetilde{Y}}_{21}}}{{%
{\widetilde{Y}}_{22}}}}|\,\,{\rm and}\,\,|{\widetilde{Y}}_{12}|=|{\widetilde{%
Y}}_{21}|.  \label{tz111}
\end{equation}
This is the (1,1) texture zero condition together with the symmetry
needed to obtain eq(\ref{eq:vub1},\ref{eq:vus1}). To proceed further
we need to determine the mixing angles in terms of the Yukawa
couplings. To do this we assume the off diagonal elements are small
relative to the on-diagonal ones in each step of the diagonalisation,
leading to the
perturbative relation for the small mixing angles given by\footnote{%
Actually this equation defines just how small the off diagonal elements need
be since successive terms in the expansion should be well ordered.}
\begin{eqnarray}
s_{23}^{Y} &\simeq &{\frac{Y_{23}}{Y_{33}}}+{\frac{{Y_{32}Y_{22}}}{Y_{33}^{2}%
}},\;s^{\prime }{}_{23}^{Y}\simeq {\frac{Y_{32}}{Y_{33}}}+{\frac{{%
Y_{23}Y_{22}}}{Y_{33}^{2}}}  \nonumber \\
s_{13}^{Y} &\simeq &{\frac{{{\widetilde{Y}}_{13}}}{Y_{33}}}+{\frac{{{%
\widetilde{Y}}_{31}Y_{11}}}{Y_{33}^{2}}},\;s^{\prime }{}_{13}^{Y}\simeq {%
\frac{{{\widetilde{Y}}_{31}}}{Y_{33}}}+{\frac{{{\widetilde{Y}}_{13}Y_{11}}}{%
Y_{33}^{2}}}  \nonumber \\
s_{12}^{Y} &\simeq &{\frac{{{\widetilde{Y}}_{12}}}{\widetilde{Y}_{22}}}
+{\frac{{{%
\widetilde{Y}}_{21}Y_{11}}}{\widetilde{Y}_{22}^{2}}},
\;s^{\prime }{}_{12}^{Y}\simeq {%
\frac{{{\widetilde{Y}}_{21}}}{\widetilde{Y}_{22}}}
+{\frac{{{\widetilde{Y}}_{12}Y_{11}}}{\widetilde{Y}_{22}^{2}}}   
\label{angles}
\end{eqnarray}
The successive rotations produce elements 
\begin{equation}
{\widetilde{\widetilde{Y}}}_{11}\simeq {\widetilde{Y}}_{11}-{\frac{{{%
\widetilde{Y}}_{12}{\widetilde{Y}}_{21}}}{{{\widetilde{Y}}_{22}}}},\;{%
\widetilde{Y}}_{11}\simeq Y_{11}-{\frac{{{\widetilde{Y}}_{13}{\widetilde{Y}}%
_{31}}}{Y_{33}}},\;{\widetilde{Y}}_{22}\simeq Y_{22}-{\frac{{Y_{23}Y_{32}}}{%
Y_{33}}},
\end{equation}
and 
\begin{equation}
\label{tilde12}
{\widetilde{Y}}_{12}=Y_{12}-Y_{13}s^{\prime }{}_{23}^{Y},\;{\widetilde{Y}}%
_{21}=Y_{21}-Y_{31}s_{23}^{Y},
\end{equation}
\begin{equation}
\label{tilde13}
{\widetilde{Y}}_{13}=Y_{13}+Y_{12}s^{\prime }{}_{23}^{Y},\;{\widetilde{Y}}%
_{31}=Y_{31}+Y_{21}s_{23}^{Y}.
\end{equation}
From this equation one may see that the contribution of terms
involving elements below the diagonal are suppressed by inverse powers
of the heavier quark masses. For this reason they are only weakly
constrained by the $CKM$ mixing angles, with the only possible
exception of the (3,2) element~\cite{Romanino:2000dz}. 
Returning to the condition
eq(\ref{tz13}) for the (1,3) texture zero prediction we see that
$Y_{13}$ must be small. The condition on $Y_{31}$ is much weaker due
to the heavy quark suppression. As we shall discuss the larger
hierarchy in the up quark masses means that the down quark
contribution to the mixing angles dominates.  Thus, to a good
approximation the requirement for the (1,3) texture zero prediction is
\begin{equation}
\tilde D_{13} \ll \frac{\tilde U_{12} D_{23}}{\tilde U_{22}}
\label{tzcond} 
\end{equation}
We are interested in what happens if this condition is {\it not} satisfied.
In this case, from eqs(\ref{vhr}) and (\ref{tz13}), we have 
\begin{eqnarray}
{\frac{|V_{ub}|}{|V_{cb}|}} &\approx &\left| \sqrt{\frac{m_{u}}{m_{c}}}-%
\frac{s_{13}}{s_{23}}\right|  \nonumber \\
{\frac{|V_{td}|}{|V_{ts}|}} &\approx &\left| \sqrt{\frac{m_{d}}{m_{s}}}-%
\frac{s_{13}}{s_{23}}\right|  \label{tap}
\end{eqnarray}
How is this analysis affected if $Y_{ij}$ are complex? We discuss this in
detail in the next Section but the implications are easy to anticipate. The
sequence of rotations in eq(\ref{rot}) will now be interspersed with various
diagonal rephasing matrices. This will change the above equations
introducing phases in various terms but cannot induce any new terms in $%
V_{ij}$. However, it is clear that even in this case eqs(\ref{tz13}) and (%
\ref{tz111}) {\em are} the correct conditions for yielding the predictions
eq(\ref{eq:vub1}). In turn this means that eq(\ref{tap}) remains correct
although the terms proportional to $\frac{s_{13}}{s_{23}}$ may acquire
different phases in the two equations (see below). However this does not
change the conclusion about which texture zero prediction receives the
dominant correction.

Note that the texture zero predictions are modified in a definite way in
that the prediction for $|V_{ub}|/|V_{cb}|$ in general has a 
larger percentage change than that for $|V_{td}|/|V_{ts}|$ because
the mass hierarchy for the up quarks is larger than that for the down quarks
while the correction proportional to $s_{13}/s_{23}$ remains the
same in both cases. This is just what is needed to correct the disagreement
we found when comparing experiment with the texture zero prediction. The
correction term to the ratio $|V_{ub}|/|V_{cb}|$ is 
\begin{equation}
\label{phU}
1-c_{U}\sin \psi +\frac{1}{2}c_{U}^{2} 
\end{equation}
where $c_{U}=\frac{s_{13}}{s_{23}}/\sqrt{\frac{m_{u}}{m_{c}}}$ and
$\psi $ is the relative phase between the two terms when the $CP$
phase
angle $\phi $ is $90^{0}$ while the correction term to the ratio $%
|V_{td}|/|V_{ts}|$ is 
\begin{equation}
\label{phD}
1-c_{D}\cos \psi +\frac{1}{2}c_{D}^{2} 
\end{equation}
where $c_{D}=\frac{s_{13}}{s_{23}}/\sqrt{\frac{m_{d}}{m_{s}}}$.  If we
take $\frac{|D_{13}|}{|D_{23}|}\approx 0.04$ and $\psi \approx
-45^{0}$ then we can easily get a 90\% increase, for example, to
$|V_{ub}|/|V_{cb}|$ while affecting $|V_{td}|/|V_{ts}|$ by only about
10\%.

\subsection{Inclusion of phases - General parametrization \label{general}}

Consider the Yukawa matrices for the case where $U_{11}$ and $D_{11}$
are both zero, i.e. just one texture zero in each. From the arguments of
Kosenko and Shrock \cite{ks} there would be then be a total of eight
unremovable phases. We may assign the eight phases as $\phi _{12}^{U,D}$, $%
\phi _{13}^{U,D}$, $\phi _{22}^{U,D}$ and $\phi _{23}^{U,D}$ so that all
possible quantities which are invariant under re-phasing transformations,
e.g. $U_{12}D_{23}U_{22}^{\ast}D_{13}^{\ast}$ are not, in
general, real -- in accordance with the discussion of Kosenko and Shrock.
Having so many phases is unnecessary as far as the physics is concerned
since the CKM matrix elements computed from eq(\ref{vhr}) in the heavy quark
limit depend on only two independent combinations of the eight phases. Using
eqs(13-16) we have 
\begin{equation}
\label{us}
V_{us}=\frac{|\tilde D_{12}|}{|\tilde D_{22}|}
e^{i(\phi _{12}^{D}-\phi _{22}^{D})}-%
\frac{|\tilde U_{12}|}{|\tilde U_{22}|}e^{i(\phi _{12}^{U}-\phi _{22}^{U})}
\end{equation}
i.e. $|V_{us}|$ depends on the phase $\phi _{1}=(\phi _{12}^{U}-\phi
_{22}^{U})-(\phi _{12}^{D}-\phi _{22}^{D})$. In addition we have 
\begin{eqnarray}
V_{ub} &=&\frac{|\tilde D_{13}|}{|\tilde D_{33}|}e^{i\phi _{13}^{D}}-\frac{%
|\tilde U_{12}|}{|\tilde U_{22}|}e^{i(\phi _{12}^{U}-\phi _{22}^{U})}\frac{%
|\tilde D_{23}|}{|\tilde D_{33}|}e^{i\phi _{23}^{D}}  \nonumber \\
V_{td} &=&-\frac{|\tilde D_{13}|}{|\tilde D_{33}|}e^{i\phi _{13}^{D}}+\frac{%
|\tilde D_{12}|}{|\tilde D_{22}|}e^{i(\phi _{12}^{D}-\phi _{22}^{D})}\frac{%
|\tilde D_{23}|}{|\tilde D_{33}|}e^{i\phi _{23}^{D}}
\label{td}
\end{eqnarray}
Thus the magnitudes $|V_{ub}|$, $|V_{td}|$ depend on only the combinations $%
\phi _{2}=(\phi _{13}^{D}-\phi _{23}^{D})-(\phi _{12}^{D}-\phi _{22}^{D})$
and $\phi _{1}-\phi _{2}$ respectively. As a result so must the Wolfenstein
parameters $\rho $ and $\eta $ depend only on the two phases $\phi _{1}$ and 
$\phi _{2}$. Moreover, evaluating the invariant 
$J={\rm Im}\{V_{cb}V_{us}V_{cs}^{%
\ast }V_{ub}^{\ast }\}$ which determines the magnitude of $CP$ violation, we
find 
\begin{equation}
\eta \propto {\rm Im}[J]=\frac{|\tilde D_{23}|}{|\tilde D_{33}|}\left[ \frac{%
|\tilde D_{23}|}{|\tilde D_{33}|}\frac{|\tilde D_{12}|}{|\tilde D_{22}|}\frac{%
|\tilde U_{12}|}{|\tilde U_{22}|}\sin \phi _{1}
-\frac{|\tilde D_{13}|}{|\tilde D_{33}|}%
\left( \frac{|\tilde D_{12}|}{|\tilde D_{22}|}\sin \phi _{2}
+\frac{|\tilde D_{12}|}{%
|\tilde U_{22}|}\sin (\phi _{1}-\phi _{2})\right) \right]  \label{imj}
\end{equation}
For small $|D_{13}|$, the first term is the leading one and if the sub-leading
corrections were negligible, then 
the $CP$ violating phase $\phi _{CP}$ would be just $\phi_{1}$, the 
{\it same} phase which enters in eq({\ref{eq:vus1}). In this case the phase 
$\phi_1$ is simply related
to the `standard' (i.e. PDG convention) of the $CP$-violating phase,
$\delta$ by $\delta = \pi - \phi_1 - \beta$ where $\beta$ is the angle 
appearing in the unitarity triangle. 
The next leading correction is the second term, proportional to 
$\sin \phi _{2}$. 
Our proof that the phase $\phi_1$ which drives the $CP$-violating phase
is the same one that appears in eq({\ref{eq:vus1}) follows
simply from the suppression of terms in the heavy quark limit and the
equality of the $(1,2)$ and $(2,1)$ elements. This generalizes the previous
proof of this result \cite{fritzsch1, hall1} which assumed an Hermitian form
for the mass matrices with texture zeros in the $(1,1)$ and $(1,3)$, $(3,1)$
elements. Although any matrix can be made Hermitian by phase changes in
general the texture zeros are not preserved by such transformations so this
is the most general starting point.

To summarize we have shown that it is sufficient in a parameterization of
the mass matrices to retain two non-zero phases which we take to be $\phi
_{12}^{U}$ and $\phi _{13}^{D}$, i.e. $\phi _{1}=\phi _{12}^{U}$ and $\phi
_{2}=\phi _{13}^{D}$. Our analysis requires $\phi _{12}^{U}\approx 90^{0}$
which is the case of `maximal' $CP$ violation for fixed quark mass ratios ($%
c.f.$ \cite{fritzsch1}).

\subsection{Fit to the data: symmetric texture}
\label{sec:u1}

We now turn to a fit to the data with a non-zero entry in the (1,3)
position of the down quark mass matrix. As we have discussed the fit
to $CKM$ matrix elements is in this case insensitive to the matrix
elements below the diagonal.  For definiteness we perform a fit making
the assumption that the mass matrix is symmetric. This corresponds to
a specific choice for the elements below the diagonal consistent with
this ``smallness'' criterion.
Of course, assuming an Hermitian rather than
a symmetric form for the mass matrices does not change the quality of
the fit.

Leaving aside a discussion of phases for the moment, the $U$ and $D$ Yukawa
matrices have the form 
\begin{equation}
U/h_{t}=\left( 
\begin{array}{ccc}
0 & b^{\prime }\epsilon ^{3} & c^{\prime }\epsilon ^{4} \\ 
b^{\prime }\epsilon ^{3} & \epsilon ^{2} & a^{\prime }\epsilon ^{2} \\ 
c^{\prime }\epsilon ^{4} & a^{\prime }\epsilon ^{2} & 1
\end{array}
\right)  \label{gpu}
\end{equation}
\label{u1}

and 
\begin{equation}
D/h_{b}=\left( 
\begin{array}{ccc}
0 & b\bar{\epsilon}^{3} & c\bar{\epsilon}^{4} \\ 
b\bar{\epsilon}^{3} & \bar{\epsilon}^{2} & a\bar{\epsilon}^{2} \\ 
c\bar{\epsilon}^{4} & a\bar{\epsilon}^{2} & 1
\end{array}
\right)  \label{gpd}
\end{equation}
\label{d1}

\noindent
where $h_{t}$, $h_{b}$ are the $t,b$ Yukawa couplings and the expansion
parameters $\epsilon $ and $\overline{\epsilon }$ will be chosen so that the
remaining parameters $a,$ $b,$ $c,$ $a^{\prime },$ $b^{\prime }$ and $%
c^{\prime }$ are all of $O(1)$. This texture is similar to the {\it ansatz}
considered by Branco et al.\cite{beg}.

In fact we can extend this parameterization to include charged lepton
masses by choosing the same matrix $L$ (with the same parameters) for
the charged leptons mass matrix as for the down quarks except for the
usual Georgi-Jarlskog \cite{gj} factor of $-3$ multiplying the (2,2)
element. This retains $m_{b}$-$m_{\tau }$ unification and also gives
good predictions for the lighter generations of lepton.

The small expansion parameters are determined immediately, since to leading
order (all our discussion is to leading order) we have 
\begin{equation}
\frac{m_{c}}{m_{t}}=\epsilon ^{2}\quad \quad \quad \quad \frac{m_{s}}{m_{b}}=%
\bar{\epsilon}^{2}  \label{eq:eps1}
\end{equation}
Since the above texture should apply at the unification scale, we have 
\begin{equation}
\epsilon \simeq 0.05\quad \quad \quad \quad \quad \bar{\epsilon}\simeq 0.15
\label{eq:epsvalues}
\end{equation}
i.e. we see that $\epsilon <\bar{\epsilon}$ and suggests $\epsilon ={\cal O}(%
\bar{\epsilon}^{2})$. The coefficients $b$ and $b^{\prime }$ are also
determined to leading order since 
\begin{equation}
\frac{m_{u}}{m_{c}}=(b^{\prime }\epsilon )^{2}\quad \quad \quad \quad \frac{%
m_{d}}{m_{s}}=(b\bar{\epsilon})^{2}  \label{eq:eps2}
\end{equation}
giving $b\simeq 1.5$ and $b^{\prime }\simeq 1$. So far the parameters are
taken real but we now introduce phases as discussed in Section \ref{general}%
. For the case of interest with a small (1,3) element only two phases play a
role in determining the physics, as discussed above. Here we assign a phase
to each of $U$ and $D$ by taking 
\begin{equation}
b^{\prime }\rightarrow b^{\prime }\;e^{i\phi
}\;\;\;\;\;\;\;\;\;\;\;\;c\rightarrow c\;e^{i\psi }  \label{eq:phase1}
\end{equation}
Having chosen to attach a phase $\phi $ to the (1,2) element of $U,$ as
discussed in section \ref{general} the one phase chosen in $D$ should be
attached to a different element of $D$. We then expand all the elements of
the rotation matrices which diagonalize $U$ and $D$ in terms of $\epsilon $
and $\bar{\epsilon}$, retaining the leading terms only. The CKM elements are
expressed in terms of these rotation matrix elements (see \cite{hall} for
example) which then allow leading order expressions for the $V_{ij}^{CKM}$
in terms of our expansion parameters.

At this leading order, we have
\begin{equation}
V_{us}=b\bar{\epsilon}-b^{\prime }\epsilon \;e^{i\phi }\;\;\Longrightarrow
|V_{us}|=\lambda =\sqrt{\frac{m_{d}}{m_{s}}}\;(1+{\cal O}(\epsilon /\bar{%
\epsilon}))  \label{eq:vus}
\end{equation}
where $\lambda $ is the Wolfenstein parameter. The Wolfenstein parameter $A$
fixes the value of $|V_{cb}|$ and in our expansion 
\begin{equation}
|V_{cb}|=A\lambda ^{2}=a\epsb^{2}+{\cal O}(\epsilon \bar{\epsilon}%
^{4})\Longrightarrow A=\frac{a}{b^{2}}\; .  \label{eq:vcb1}
\end{equation}
We are neglecting here the up quark contribution to $|V_{cb}|$, which
is justified if the (2,3) elements in the up quark matrix are indeed
of order $\epsilon^2$ as suggested by eq.~(\ref{gpu}). From a
phenomenological point of view, however, a larger size for those
elements (e.g. of order $\epsilon$) is also allowed and could lead to
non negligible up quark contributions to $V_{cb}$.

Since the textures that we are discussing apply at the unification scale
rather than at low energies, we must use values for mass ratios and CKM
parameters appropriate to that scale. For this it is convenient to introduce
the parameter $\chi =(M_{X}/M_{Z})^{-h_{t}^{2}/(16\pi ^{2})}\approx 0.7$ and
then 
\begin{eqnarray}
\frac{A(M_{X})}{A(M_{Z})} &=&\chi  \nonumber \\
\frac{(m_{s}/m_{b})(M_{X})}{(m_{s}/m_{b})(M_{Z})} &=&\chi  \nonumber \\
\frac{(m_{c}/m_{t})(M_{X})}{(m_{c}/m_{t})(M_{Z})} &=&\chi ^{3}
\end{eqnarray}
Thus at the unification scale we have $A\simeq 0.58$ or $a\simeq 1.3$
As discussed above, up to corrections suppressed by inverse powers of
the third generation masses, the phase $\phi $ determines the sign and
magnitude of the CP-violating CKM phase. In the SM context, the
observed CP\ violation requires a near maximal phase, $\phi \approx
90^{0},$ so
\begin{eqnarray}
V_{ub} &=&c\epsb^{4}e^{i\psi }-iab^{\prime }\epsilon \bar{\epsilon}^{2} 
\nonumber \\
V_{td} &=&-c\bar{\epsilon}^{4}e^{i\psi }+ab\bar{\epsilon}^{3}
\label{eq:vubvtd}
\end{eqnarray}
which imply that, to leading order the Wolfenstein parameters which govern
the size of $V_{ub}$ and $V_{td}$ are given by 
\begin{eqnarray}
\rho &=& \left( \frac{b^{\prime}\eps}{b\epsb }\right)^2
 + \frac{\epsb}{ab}\; c\cos\psi 
 - \frac{b^{\prime}\eps}{ab^2}\; c\sin \psi \nonumber \\
\eta &=& \;\;\; \frac{b^{\prime }\eps}{b\epsb}
 \;\; - \;\frac{\epsb}{ab}\; c\sin \psi 
 - \frac{b'\eps}{ab^2}\; c\cos \psi
\label{eq:rhoeta}
\end{eqnarray}
For $c$ small, the phase $\phi \simeq +90^{0}$ fixes the correct 
sign of the first (dominant) term in
the expression for $\eta $ and maximizes $CP$ violation for fixed quark mass
ratios. In passing we note that, to this order, the entire list of quark
masses and CKM matrix elements do not involve $a^{\prime }$ or $c^{\prime }$
and so could take on any value of $O(1)$ without affecting the physics. We
can express the perturbation to the canonical values for $|V_{ub}/V_{cb}|$
and $|V_{td}/V_{ts}|$ given by eq(\ref{eq:vub1}) 
\begin{eqnarray}
\frac{|V_{ub}|^2}{|V_{cb}|^2} &=&\frac{m_{u}}{m_{c}}\;\left( 1-%
2\frac{c\bar{\epsilon}^{2}\sin \psi }{ab^{\prime }\epsilon }+\frac{c^{2}\bar{%
\epsilon}^{4}}{a^{2}b^{\prime }{}^{2}\epsilon ^{2}}\right)  \nonumber \\
\frac{|V_{td}|^2}{|V_{ts}|^2} &=&\frac{m_{d}}{m_{s}}\;\left( 1-%
2\frac{c\bar{\epsilon}\cos \psi }{ab} + \frac{c^2\epsb^2}{a^2b^2}
\right)  \label{eq:vub2}
\end{eqnarray}
The effect of filling in
the (1,3) texture zero of $D$ can be more dramatic for $|V_{ub}/V_{cb}|$
than for $|V_{td}/V_{ts}|$ since the correction to the latter is suppressed
by $\eps/\epsb \sim \sqrt{m_u/m_c}/\sqrt{m_d/m_s}$ relative to the
correction to the former. We can get the desired
phenomenological result of moving $|V_{ub}/V_{cb}|$ up towards the measured
value around 0.09 while not unduly perturbing 
the value of $|V_{td}/V_{ts}|$ given by
eq(\ref{eq:vub1}).

\begin{figure}[ht]
\vspace*{-0.75cm} 
\begin{center}
\mbox{\epsfig{file=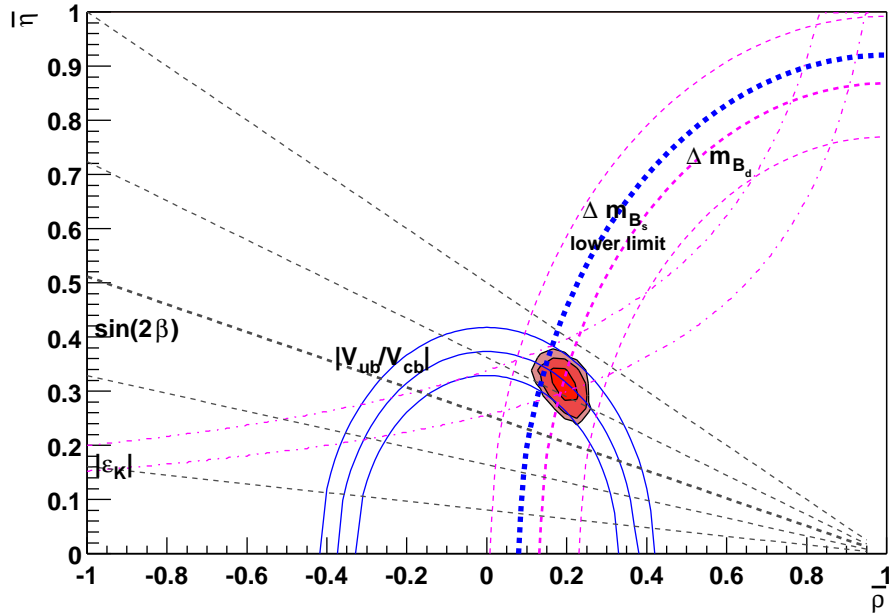,height=9cm}}
\end{center}
\vspace*{-0.75cm}
\caption{\small{Fit A of Section 3.2 (symmetric texture) to the
measurements of $|V_{ub}/V_{cb}|$, $\Delta m_{B_s}$, $\Delta m_{B_d}$,
$|\epsilon_K|$ and $\sin 2\beta$.}}
\label{fig:fclst2o3}
\end{figure}

\begin{figure}[ht]
\vspace*{-0.75cm} 
\begin{center}
\mbox{\epsfig{file=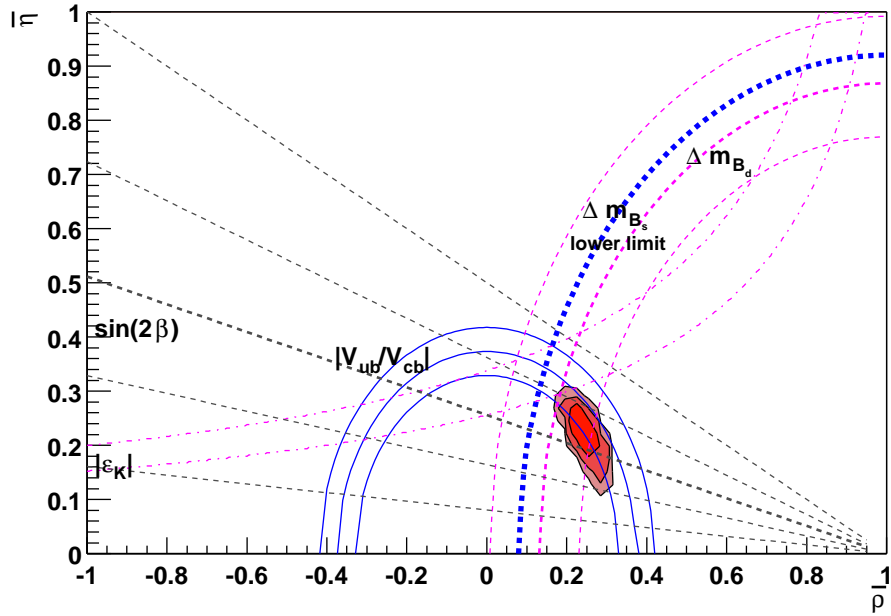,height=9cm}}
\end{center}
\vspace*{-0.75cm}
\caption{\small{Fit A of section 3.2 (symmetric texture) to the
measurements of $|V_{ub}/V_{cb}|$ and $\sin 2\beta$. }}
\label{fig:fclst2o5}
\end{figure}

\begin{table}[th]
\label{massesQ}
\par
\begin{center}
\begin{tabular}{|c|c|c|}
\hline
Parameter & Value & Referen. \\ 
$Q$ & $22.7\pm 0.8$ & \cite{Leutwyler:2000hx} \\ 
$m_{u}/m_{d}$ & $0.533\pm 0.043$ & \cite{Leutwyler:2000hx} \\ 
$m_{c}/m_{s}$ & $9.5\pm 1.7$ & \cite{pdgb:dat} \\ \hline
\end{tabular}
\end{center}
\caption{Values for the quark masses ratios and the parameter $Q$ 
(defined in eq(\protect\ref{eq:qq})) used for the texture zero fits.}
\end{table}

Using the expansions of eqs({\ref{gpu},\ref{gpd}) for the $U$ and $D$
matrices in terms of $\epsilon $ and $\bar{\epsilon}$ we carry out a fit
using information on the measured estimates for ratios of quark masses and
CKM matrix elements. The expansion parameters $\epsilon $, $\bar{\epsilon}$
and the ${\cal O}(1)$ coefficients $a$, $b$, $b^{\prime }$ and $c$ are
determined via the expressions (\ref{eq:eps1},\ref{eq:eps2},\ref{eq:vus}%
,\ref{eq:vcb1},\ref{eq:rhoeta},\ref{eq:vub2}). We follow BHR in using the
combination $Q$ given by 
\begin{equation}
Q=\frac{m_{s}/m_{d}}{\sqrt{1-(m_{u}/m_{d})^{2}}}  \label{eq:qq}
\end{equation}
which is determined accurately from chiral perturbation theory.
Additionally we can use the ratios of the masses $m_{u}/m_{d}$,
$m_{c}/m_{s}$.  

The resulting fit (Fit A) yields the values
\begin{eqnarray}
\bar{\epsilon} &=&0.15\pm 0.01\quad |b|=1.5\pm 0.1\quad a=1.31\pm 0.14 
\nonumber \\
\centering{|c|} &=&{2.7\pm 0.10\quad \psi =-24^{0}\pm 3^{0}}
\end{eqnarray}
where the values are given at the unification scale.

Demanding a texture zero in the (1,3),(3,1) elements for a symmetric $D$
matrix leads, in particular,
to too small a value for $\bar{\rho}$ and there is a marked improvement in
the overall description of the data when this zero is filled in, as can be
seen by comparing Fig.~\ref{fig:fclst2o3} with Fig.~\ref{fig:fclst1}. 

That $c \sim 3$ means that the order of the (1,3) term is ambiguous and
could be either {\cal O}($\bar{\epsilon}^{4})$ or 
{\cal O}($\bar{\epsilon}^{3})$. \footnote{%
That $c^{\prime }$ is quite undetermined means that a texture zero in $%
U_{13} $, $U_{31}$ is not ruled out.} While the texture zero in $D_{13}$, $%
D_{31}$ does lead to the rather attractive result of eq(\ref{eq:vub1}) the
current experimental data, within the context of the Standard Model, favour
perturbative corrections as suggested by eqs(\ref{eq:vub2}).

We noted earlier (Section~\ref{SMfit}) that there is a potential
disagreement in the SM fit to $|\epsilon_K|$, $\Delta m_{B_s}$, 
$\Delta m_{B_d}$ and the recent measurement of 
$\sin 2\beta = 0.41\pm0.17$\cite{BB:2001}. 
To quantify
the implications of this, we perform a second version of Fit A (still using the 
texture of eqs({\ref{gpu},\ref{gpd})),
dropping the constraints
of $|\epsilon_K|$, $\Delta m_{B_s}$ and $\Delta m_{B_d}$. This is shown in 
Fig.~\ref{fig:fclst2o5}. The effect is to increase slightly the value of
the parmeter $c$ to $3.31\pm0.10$ and the value of $\psi$ to $6^{0}\pm3^{0}$. 

In Fig.~\ref{fig:f1dexpvst2} we show the resulting 
probability distributions for the quantities $\frac{V_{ub}}{V_{cb}}$,
$|\epsilon_K|$, $\Delta m_{B_s}$ and $\sin 2\beta$ compared to the 
corresponding experimental distributions.

There is actually a further solution (Fit B) to the above equations where 
the phase $\phi \sim -90^0$ rather than $+90^0$ which corresponds to 
to $b^{\prime} \rightarrow -b^{\prime}$ in 
eqs(\ref{eq:rhoeta},\ref{eq:vub2}). 
In this case the
first term in the expression for $\eta$ is no longer the dominant one and
a larger value of $c$ is needed to obtain a positive value for $\eta$. 
Only the parameters $c$ and $\psi$ change from Fit A, and for Fit B we
find $c = 8.45 \pm 0.33$, $\psi = -58^0 \pm 5^0$. The result of Fit B is shown 
in Fig.~\ref{fig:fclst2o4} and comparing with Fig.~\ref{fig:fclst2o3} we
see no difference in the quality of fits A and B.
This solution has the (1,3) matrix element of the same order
({\cal O}($\epsb^3$)) as the (1,2) matrix element. In this case
$D_{13}/D_{12} \approx D_{23}/D_{22} \approx 1$, suggestive of a non-Abelian 
family structure.

\begin{figure}[ht] 
\vspace*{-0.75cm}
\begin{center}
\mbox{\epsfig{file=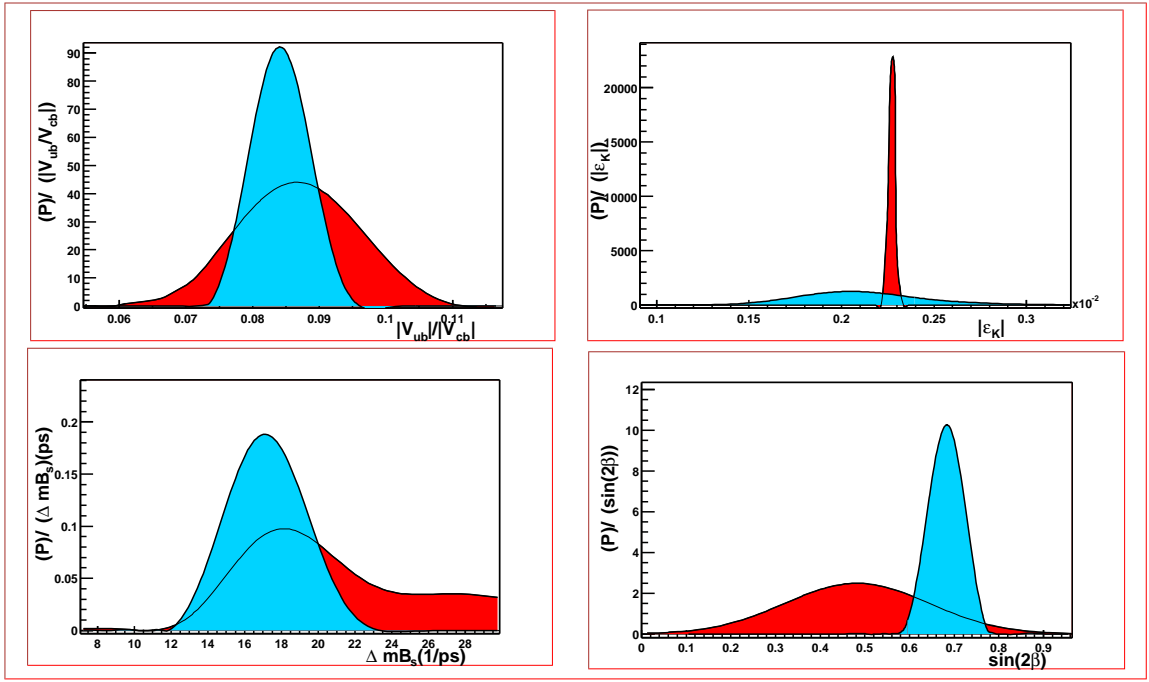,height=9cm}}
\end{center}
\vspace*{-0.75cm}
\caption{\small{One dimensional probabilities for the physical observables
$|V_{ub}|/|V_{cb}|$, $|\epsilon_K|$, $\Delta m_{B_d}$ and $\sin 2\beta$. 
The probabilities in red (dark) correspond to the experimental constraints 
and the probabilities in blue (light) correspond to the predictions of the 
texture with 13 and 31 entries different from zero with $c=(2.7\pm 0.10)$ 
and $\psi=(-24\pm 3)^{0}$.}}
\label{fig:f1dexpvst2}
\end{figure}

\begin{figure}[ht]
\vspace*{-0.75cm} 
\begin{center}
\mbox{\epsfig{file=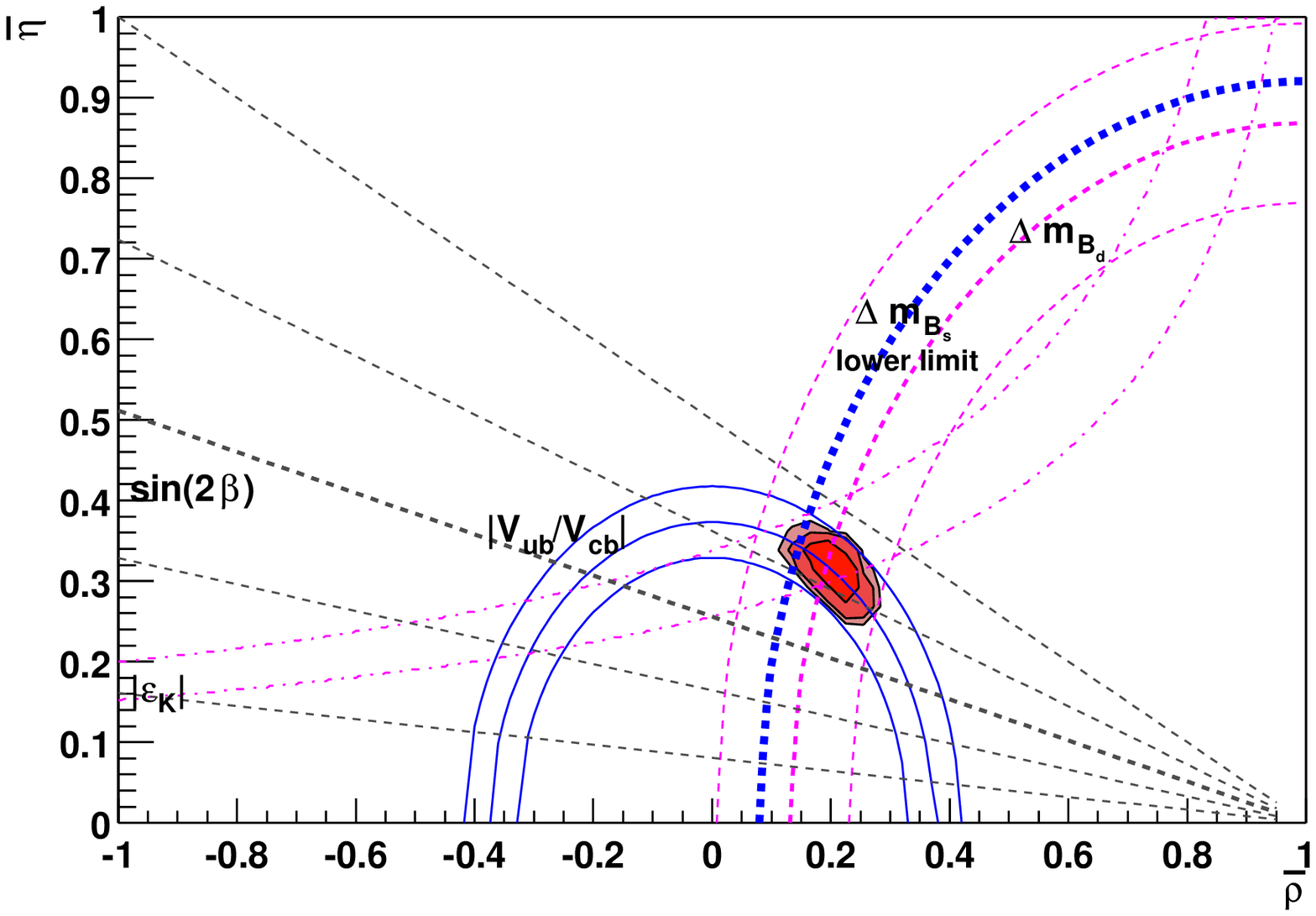,height=9cm}}
\end{center}
\vspace*{-0.75cm}
\caption{\small{Fit B of Section 3.2 (symmetric texture) to the
measurements of $|V_{ub}/V_{cb}|$, $\Delta m_{B_s}$, $\Delta m_{B_d}$,
$|\epsilon_K|$ and $\sin 2\beta$.}}
\label{fig:fclst2o4}
\end{figure}

\subsection{Fit to the data: asymmetric texture} 
\label{sec:u2}

We now consider a fit to the data in which the texture zero relation
$|V_{ub}/V_{cb}| = \sqrt{m_u/m_c}$ is modified by higher order
corrections in the perturbative diagonalization of the down mass
matrix. As in the previous section, the correction to that relation
comes from a small but non negligible rotation $s^D_{13}$ induced by a
non zero element $\tilde D_{13}$.  Here, however, we assume that
$\tilde D_{13}$ is mainly induced by the rotation $\srdd$ used to
diagonalize the 23 sector of the down quark mass matrix, the initial
value $D_{13}$ being negligible (see eq.~(\ref{tilde13})).  This
situation is therefore complementary to the one considered in the
previous subsection, where $\tilde D_{13}$ was mainly given by the
original entry $D_{13}$ and the contribution to $\tilde D_{13}$
proportional to $\srdd$ was assumed to be negligible.

Let us first of all estimate how large $\srdd$ should be in order to give a
significant contribution to $|V_{ub}/V_{cb}|$. We start from
eqs.~(\ref{tap}), that assume only $|Y_{12}|=|Y_{21}|$,
$Y_{11}=0$. The size of the correction is determined by $s^Y_{13}$,
which can be written as
\begin{equation}
  \label{s13}
  s^Y_{13} = \frac{\tilde Y_{13}}{Y_{33}} \sim
  \sqrt{\frac{m_1}{m_2}}\frac{m_2}{m_3} s^{\prime Y}_{23} \;,
\end{equation}
where $m_i$ is the mass of the quark of the $i$-${th}$ family in the
sector $Y=U,D$ and we have used $\tilde Y_{13} = Y_{12} s^{\prime
  Y}_{23}$, $Y_{12}/\tilde Y_{22} \sim \sqrt{m_1/m_2}$ and $\tilde
Y_{22}/Y_{33} \simeq m_2/m_3$. From eq.~(\ref{s13}) one can see
that the contribution of the up quark rotation $s^U_{13}$ to
$s_{13}=s^D_{13}-s^U_{13}$ is negligible. The mass ratios in
eq.~(\ref{s13}) are in fact much smaller in the up sector than in the
down sector. Moreover, the factor $s^{\prime Y}_{23}$ in~(\ref{s13})
appears in the product $s^{\prime Y}_{23}s^{Y}_{23}$, which
contributes to the ratio $m_2/m_3$. Barring cancellations, we therefore
have $s^{\prime Y}_{23}s^{Y}_{23}\circa{<}m_2/m_3$, a constraint
stronger in the up sector due again to $m_c/m_t\ll m_s/m_b$. Hence we can
safely neglect $s^U_{13}$ and write
\begin{equation}
  \label{ubcorr}
  \frac{s_{13}}{s_{23}} \simeq \frac{s^D_{13}}{s_{23}} \sim
  \sqrt{\frac{m_d}{m_s}}\frac{m_s/m_b}{|V_{cb}|} s^{\prime D}_{23} \; .
\end{equation}
Since at the electroweak scale we have $\sqrt{m_d/m_s}\sim 0.22$ and
$(m_s/m_b)/|V_{cb}|\sim 0.6$, in order to get a correction
$s_{13}/s_{23}\sim 0.03$ (which gives a good agreement with data in
absence of phases) it is sufficient to have $\srdd \sim 0.2 - 0.3$, or
$|D_{32}| \sim (0.2 - 0.3) |D_{33}|$. Larger values are also in
principle allowed depending on the phases in eq.~(\ref{phU}).
Notice that in first approximation the ratio $D_{32}/D_{33}$ at the
unification scale is the same as at the electroweak scale.

Quark mass textures with $|D_{32}/D_{33}| = \ord{1}$ have been
considered in the literature~\cite{Albright:1998vf} in connection with a
large leptonic mixing angle originating from the charged lepton mass
matrix. In SU(5) unified models, the ratio $|D_{32}/D_{33}|$ associated
with the right-handed quark rotation $\srd$ corresponds in fact to the
ratio of charged lepton mass matrix elements $|E_{23}/E_{33}|$
associated with a left-handed charged lepton rotation that mixes $\mu$
and $\tau$ neutrinos. On one hand, this link can be considered as a
motivation for studying asymmetric textures as a solution of the
$V_{ub}/V_{cb}$ problem. From the opposite point of view, we can say
that in the context of unified models, the asymmetric solution of the
$V_{ub}/V_{cb}$ problem, pointing at a largish value of
$|D_{32}/D_{33}|$, indicates that the large mixing angle responsible for
the atmospheric neutrino anomaly comes, or at least receives a
significant contribution, from the diagonalization of the charged lepton
mass
matrix.

The precise relation between the right-handed quark and left-handed
lepton rotations depends on possible Clebsh coefficients relating
transposed down quark and charged lepton matrix elements. The presence
of
non-trivial coefficients enhancing some charged lepton matrix element is
indeed suggested by the empirical relation $m_\mu/m_\tau \sim 3
m_s/m_b$. Since a value of $|D_{32}/D_{33}|$ around 1/3 is preferred by
a fit of data, a near to maximal lepton mixing can be obtained if the
Georgi-Jarskog factor 3 sits in the $E_{23}$ entry.

Notice that a sizeable $|D_{32}|\sim |D_{33}|/3$ indicates an
asymmetry $|D_{32}|\gg|D_{23}|$. The ratio $|D_{23}/D_{33}|$ is in
fact expected to be of order $|V_{cb}|\ll 1/3$, barring cancellations
between up and down quark contributions to $V_{cb}$. Such an asymmetry
can be easily obtained in the context of Abelian and non-Abelian
models. In Section~\ref{sec:u2model} we will describe an explicit
example of non-Abelian family symmetry leading to the asymmetry in the
23 sector while preserving the relation $|Y_{12}|=|Y_{21}|$ in the 12
sector and the texture zeros $Y_{13}\simeq Y_{31}\simeq Y_{11}\simeq
0$. This texture allows us to isolate and study the corrections to the
texture zero relations we are considering in this section.  

To examine the implications of this scenario we modify the parameterisation 
of the Yukawa matrices given in eqs(\ref{gpu}, \ref{gpd}) by restoring the 
exact texture zeros in the (1,3) and (3,1) elements and allow 
$|D_{32}| \gg |D_{23}|$. 

We consider the following parameterisations for the 
mass matrices for the up and down quarks whose
absolute values can be parametrized as
\begin{eqnarray}
  \label{u2textureU}
  |U/h_t| &=&
\left( 
  \begin{array}{ccc}
    0 & c\epsilon\epsilon' & 0 \\
    c\epsilon\epsilon' & \beta \epsilon^2 & b\epsilon \\
    0 & a\epsilon & 1
  \end{array}
\right) \\
  \label{u2textureD}
  |D/h_b| &=&
\left( 
  \begin{array}{ccc}
    0 & \epsilon' & 0 \\
    \epsilon' & \alpha \epsilon & \epsilon \\
    0 & t & 1
  \end{array}
\right) \; ,
\end{eqnarray}
with $\epsilon' < \epsilon \ll 1$. 
In this parameterisation of $D$ we see, comparing with eq(\ref{gpd}),
we have changed the (3,2) element to be {\cal O}(1). The remaining elements
are of the same order, $\eps$ in eq(\ref{u2textureD}) being {\cal O}($\epsb^2$)
and $\epsilon'$ is {\cal O}($\eps^3$). The parameterisation of $U$ has the 
same form as eq(\ref{gpu}) with the exception of the (2,3)  and (3,2) elements
which are now of {\cal O}($\eps$) and not {\cal O}($\eps^2$). 
We have chosen this form to relate to a promising texture model discussed
in Section\ref{sec:u2model}. However these elements are poorly determined
by the data and it is possible to obtain solutions where $a,b$ are {\cal O}
($\eps$), corresponding to the original symmetric parametrisation of 
eq(\ref{gpu}).

In order to obtain the precise form
of the corrected texture zero relations, we first write the general
expression for $|V_{us}|$, $|V_{ub}/V_{cb}|$, $|V_{td}/V_{ts}|$ in
terms of the rotations defined by eq.~(\ref{rot}):
\begin{eqnarray}
  \label{Vus}
|V_{us}| &=& \left|\,t^D_{12}-t^U_{12} e^{i\phi_1}\right|c^D_{12}
\nonumber \\  
\left|\frac{V_{ub}}{V_{cb}}\right| &=& \left|\, t^U_{12} -\frac{s_{13}}{s_{23}}
e^{i(\phi_2-\phi_1)} \right| \label{eq:general} \\
\left|\frac{V_{td}}{V_{ts}}\right| &=& \left|\, t^D_{12} -\frac{s_{13}}{s_{23}}
e^{i\phi_2} \right| \label{RVds} \; , \nonumber
\end{eqnarray}
where $t^D_{12}$ and $t^U_{12}$ are the tangent of the 12 rotation of
left-handed down and up quark respectively. Written as above in terms
of the tangent of the angles (and cosines $c^D_{12}$, $c^U_{12}$), the
expressions are exact up to $\ord{\lambda^4}$ corrections, $\lambda$
being one of the Wolfenstein parameters. The phases $\phi_1$, $\phi_2$
were discussed in Section~\ref{general}. In order to obtain the
relation between mixing angles and quark mass ratios generalizing the
texture zero relations, one has to express the angles in
eqs.~(\ref{eq:general}) in terms of quark masses. In the case of
textures~(\ref{u2textureD},\ref{u2textureU}), we obtain
\begin{eqnarray}
t^U_{12} &=& \sqrt{\frac{m_u}{m_c}} \nonumber \\
t^D_{12} &=& \sqrt{c\frac{m_d}{m_s}}
\left(1-\frac{1}{2}t^2 c\,\frac{m_d}{m_s}\right) \label{angmas} \\
\frac{s_{13}}{s_{23}} &=& t
\sqrt{c\frac{m_d}{m_s}} \frac{m_s/m_b}{|V_{cb}|} \; , \nonumber
\end{eqnarray}
where $t=|D_{32}/D_{33}|$ as in eq.~(\ref{u2textureD}). The parameter
$t$ represents the tangent $t^{\prime D}_{23}$ of the 23 rotation on
the right-handed down quarks in eq.~(\ref{rot}). Since we are
considering the possibility of a sizeable $t$, we do not approximate
the cosine of that angle, $c\equiv 1/\sqrt{1+t^2}$, with $1$. As a
consequence we have, at leading order in $\lambda^2=\ord{m_d/m_s}$,
\[
t^D_{12}\simeq \sqrt{c \frac{m_d}{m_s}}\neq \sqrt{\frac{1}{c}
  \frac{m_d}{m_s}} \simeq t^{\prime D}_{12} \; .
\]
This is because the diagonalization of the 23 sector in the down
sector not only induces an effective (1,3) element, but also generates
a slight asymmetry in the 12 sector: $|\tilde D_{12}|\neq |\tilde D_{21}|$
despite $|D_{12}|\neq |D_{21}|$. The expression for $t^D_{12}$ above
also includes a next to leading correction in
$\lambda^2=\ord{m_d/m_s}$. 

Let us now turn to the numerical determination of the parameters
entering the expressions for $U$ and $D$. Three of those parameters,
$t$ and the two phases $\phi_1$, $\phi_2$, can be determined
independently of the values of the others in terms of $|V_{us}|$ and
$\rhob$, $\etab$ (as obtained from the SM fit). This is possible
since, for given values of the quark masses,
eqs.~(\ref{eq:general},\ref{angmas}) relate $t$, $\phi_1$, $\phi_2$ to
$|V_{us}|$ and $|V_{ub}/V_{cb}|$, $|V_{td}/V_{ts}|$, and therefore to
$|V_{us}|$ and $\rhob$, $\etab$. More precisely, to get the best
values of $t$, $\phi_1$, $\phi_2$, we use the following procedure.
First we calculate $|V_{ub}/V_{cb}|$, $|V_{td}/V_{ts}|$ in terms of
$\rhob$, $\etab$. Then, for any given value of the ratio
$t=|D_{32}/D_{33}|$ and of the phase\footnote{To be precise, in order
  to include all sign ambiguities in a single phase we actually fix
  the value of the phase $\phi'_2$ defined by
  $\cos\phi'_2=\cos\phi_2$, $\sin\phi'_2=\sin\phi_2
  \mbox{sign}(\sin\phi_1)$. In the following, $\phi_2$ should be read
  $\phi'_2$.}  $\phi_2$ we recover $m_d/m_s$ and $m_u/m_c$ from
eqs.~(\ref{eq:general},\ref{angmas}) (the phase $\phi_1$ is obtained,
up to discrete ambiguities, from the relation involving $|V_{us}|$).
Finally, we calculate $Q$, $m_u/m_d$ in terms of $m_d/m_s$, $m_u/m_c$.
We can at this point perform a fit of $Q$, $m_u/m_d$ in terms of
$\rhob$, $\etab$ for any given value of $t$ and $\phi_2$. The
quantities $m_c/m_s$, $m_s/m_b$, $|V_{cb}|$, $\lambda$ involved in the
relation between $Q$, $m_u/m_d$ and $\rhob$, $\etab$ are also included
in the fit. As mentioned before, using $Q$, $m_u/m_d$ instead of
$m_d/m_s$, $m_u/m_c$ improves considerably the quality of the fit,
especially for small $t$ (despite it requiring the inclusion of the
ratio $m_c/m_s$)~\cite{barbieri}. We then obtain $t= 0.3$,
$\phi_2\simeq -0.2\,\pi$, $\cos\phi_1\simeq 0.1$. Notice that $\phi_1$
again turns out to be almost maximal as was the case with symmetric
textures. 

The determination of $t$, $\phi_1$, $\phi_2$ described above depends
on the additional parameters in~(\ref{u2textureU},\ref{u2textureD})
only through the quark masses and $|V_{cb}|$. As a consequence, the
fit of all data in terms of all parameters decouples into a fit of
$|V_{us}|$, $\rhob$, $\etab$ in terms of $t$, $\phi_1$, $\phi_2$
(which makes use of the experimental values of the quark masses and
$|V_{cb}|$) and a fit of quark masses and $|V_{cb}|$ in terms of the
additional parameters. The first fit has been described above and is
independent of the specific form of the
textures~(\ref{u2textureD},\ref{u2textureU}) and of the values of the
additional parameters one uses to account for the quark masses and
$|V_{cb}|$. For example, independently of whether the up
quark matrix is fully symmetric or not, the parameters $a$ and $b$ are
of {\cal O}(1) or smaller, $\alpha$, $\beta$ are of {\cal O}(1) or
vanishing, provided that the values of the quark masses and of
$|V_{cb}|$ can be accounted for. All the dependence on the specific
form of~(\ref{u2textureU},\ref{u2textureD}) is therefore confined to
the fit of quark masses in terms of the parameters $\epsilon$,
$\epsilon'$, $a$, $b$, $c$, $\alpha$, $\beta$. Here we consider in
more detail such a fit in the case motivated by the flavour model
described in Section~\ref{sec:u2model} in which $\alpha = \beta = 0$.

Let us start from the quark masses. Besides $m_t$, $m_b$, trivially
accounted for by $h_t$, $h_b$, we have to account for $m_s/m_b$,
$m_c/m_t$, $m_d m_s/m^2_b$, $m_u m_c/m_t^2$. We work at the $m_t$
scale. In order to account for $m_s/m_b \simeq t\,\epsilon$ we need
$\epsilon \simeq 0.08$.  From $m_c/m_t \simeq ab\epsilon^2$ we then
have $ab\simeq 0.6$. Since $m_d m_s/m_b^2 =
{\epsilon'}^2/(1+t^2)^{3/2}$, we can obtain $\epsilon'=0.006$. The
ratio $m_u m_c/m_t^2 = (c\epsilon\epsilon')^2$ then gives $c \simeq
0.5$. Notice that all parameters that are supposed to be of {\cal O}(1)
indeed are. In particular, we have
$|U_{23}/h_t|\sim|U_{32}/h_t|\sim|D_{23}/h_b|$, which justifies the
use of the same parameter ($\epsilon$) in both the $U$ and $D$
matrices. In the model of Section~\ref{sec:u2model} the same parameter
indeed appears in those entries because the same vev generates them.
Finally, we have to account for the value of $|V_{cb}|$, which has a
contribution from the down sector, $\epsilon$, and one from the up
sector, $b\,\epsilon$. Both contributions are of the right order of
magnitude, so it is clear that $|V_{cb}|$ can be obtained for an
appropriate choice of the {\cal O}(1) coefficient $b$. The precise value
of $b$ depends on the relative phase between the two contributions.
For $\alpha=\beta=0$, the phases of the SM quark multiplets can be
redefined in such a way that $U/h_t$ and $D/h_b$ differ from their
absolute values in~(\ref{u2textureD},\ref{u2textureU}) only by a phase
$e^{i\phi}$ multiplying $t$ in $D_{32}$ and a phase $e^{i\psi}$
multiplying $b$ in $U_{23}$. The phases $\phi_1$ and $\phi_2$ are then
given by
\bea
\label{phases1}
e^{i\phi_1} &=& e^{i(\psi-\phi)} \\
\label{phases2}
e^{i\phi_2} &=& \frac{\epsilon\left(b e^{i\psi} -1\right)}{|V_{cb}|}
\; .  
\eea
The relative phase between the two contributions to $V_{cb}$ turns out
to be $e^{i\psi}$ and can be determined from the equations above.
We are actually interested to the value of $b$ which simply follows
from eq.~(\ref{phases2}): $b = |e^{i\phi_2}|V_{cb}|/\epsilon+1| \simeq
1.4$. We then also have $a\simeq 0.4$. Notice that $a$ and $b$ are
both of {\cal O}(1) but $b>a$. This slight asymmetry is reduced by the
running up to the unification scale. 

The results for this fit with the asymmetric texture, $D_{32} \gg D_{23}$ are 
very close to those for the symmetric texture of 3.2 and the resulting 
contour plot in the $\rhob$ -- $\etab$ plane is essentially identical to that
in fig.~\ref{fig:fclst2o3}.

\section{Implications for a family symmetry\label{family}}

Of course the underlying motivation for studying the detailed structure of
the quark and lepton mass matrices is that they may lead to an insight about
the structure beyond the Standard Model. Here we briefly comment on the
implications of our analysis for such structure concentrating on the
possibility there is an extension of the symmetries of the Standard Model to
include a family symmetry.

\subsection{Symmetric case}
\label{sec:famsym}

\begin{table}[tbp] \centering%
\begin{center}
\begin{tabular}{|c|cccccccc|}
\hline
& $Q_i$ & $u^c_i$ & $d^c_i$ & $L_i$ & $e^c_i$ & $\nu^c_i$ & $H_2$ & $H_1$ \\ 
\hline
$U(1)_{FD}$ & $\alpha _i$ & $\alpha _i$ & $\alpha _i$ & $a_i$ & $a_i$ & $a_i 
$ & $-2\alpha _1$ & $-2\alpha _1$ \\ \hline
\end{tabular}
\end{center}
\caption{ $U(1)_{FD}$ symmetries. } 
\label{table:2a}
\end{table}

It turns out to be remarkably easy to construct a model leading to the
mass matrices in eqs.~(\ref{gpu},\ref{gpd}) through the introduction
of an Abelian gauge symmetry, $U(1)$ (such additional symmetries
abound in string theories). The most general charge assignment of the
Standard Model states is given in Table \ref {table:2a}. This follows
since the need to preserve $SU(2)_{L}$ invariance requires
(left-handed) up and down quarks (leptons) to have the same charge.
This together with the requirement of symmetric matrices then requires
that all quarks (leptons) of the same i-th generation transform with
the same charge $\alpha _{i}(a_{i})$. If the light Higgs, $H_{2}$,
$H_{1}$, responsible for the up and down quark masses respectively
have $U(1)$ charge so that only the (3,3) renormalisable Yukawa
coupling to $H_{2}$, $H_{1}$ is allowed, only the (3,3) element of the
associated mass matrix will be non-zero as desired. The remaining
entries are generated when the $U(1)$ symmetry is broken. A
particularly interesting example may be constructed in a
supersymmetric extension of the Standard Model \cite{ir}. We assume
this
breaking is spontaneous via Standard Model singlet fields, $\theta ,\;\bar{%
\theta}$, with $U(1)_{FD}$ charge -1, +1 respectively, which acquire vacuum
expectation values (vevs), $<\theta >,\;<\bar{\theta}>$, along a ``D-flat''
direction. After this breaking all entries in the mass matrix become
non-zero. For example, the (3,2) entry in the up quark mass matrix appears
at $O(\epsilon ^{\mid \alpha _{2}-\alpha _{1}\mid })$ because U(1) charge
conservation allows only a coupling $c^{c}tH_{2}(\theta /M_{2})^{\alpha
_{2}-\alpha _{1}},\;\alpha _{2}>\alpha _{1}$ or $c^{c}tH_{2}(\bar{\theta}%
/M_{2})^{\alpha _{1}-\alpha _{2}},\;\alpha _{1}>\alpha _{2}$ and we have
defined $\epsilon =(<\theta >/M_{2})$ where $M_{2}$ is the unification mass
scale which governs the higher dimension operators. As discussed in
reference \cite{ir} one may expect a different scale, $M_{1}$ for the down
quark mass matrices (it corresponds to mixing in the $H_{2}$, $H_{1}$ sector
with $M_{2}$, $M_{1}$ the masses of heavy $H_{2}$, $H_{1}$ fields). Thus we
arrive at mass matrices of the form

\begin{eqnarray}
\frac{M_u}{m_t}\approx \left( 
\begin{array}{ccc}
h_{1 1}\rho_{11 }\epsilon_a^{\mid 2+6a \mid } & h_{1 2}\rho_{12
}\epsilon_b^{\mid 3a \mid } & h_{1 3}\rho_{13 }\epsilon_a^{\mid 1+3a\mid }
\\ 
h_{2 1}\rho_{21 }\epsilon_b^{\mid 3a \mid } & h_{2 2}\rho_{22 }\epsilon^{ 2 }
& h_{2 3}\rho_{23 }\epsilon^{ 1 } \\ 
h_{3 1}\rho_{31 }\epsilon_a^{\mid 1+3a \mid } & h_{3 2}\rho_{32 }\epsilon^{1
} & h_{3 3}
\end{array}
\right)  \label{eq:mu0}
\end{eqnarray}

\begin{equation}
\frac{M_{d}}{m_{b}}\approx \left( 
\begin{array}{ccc}
k_{11}\sigma _{11}\bar{\epsilon }_{a}^{\mid 2+6a\mid } & k_{12}\sigma _{12}%
\bar{\epsilon _{b}}^{\mid 3a\mid } & k_{13}\sigma _{13}\bar{\epsilon _{a}}%
^{\mid 1+3a\mid } \\ 
k_{21}\sigma _{21}\bar{\epsilon _{b}}^{\mid 3a\mid } & k_{22}\sigma _{22}%
\bar{\epsilon}^{2} & k_{23}\sigma _{23}\bar{\epsilon}^{1} \\ 
k_{31}\sigma _{31}\bar{\epsilon _{a}}^{\mid 1+3a\mid } & k_{32}\sigma _{32}%
\bar{\epsilon}^{1} & k_{33}
\end{array}
\right)  \label{eq:massu}
\end{equation}
where $\bar{\epsilon}=(\frac{<\theta >}{M_{1}})^{|\alpha _{2}-\alpha _{1}|}$%
, $\epsilon =(\frac{<\theta >}{M_{2}})^{|\alpha _{2}-\alpha _{1}|}$, and $%
a=(2\alpha _{1}-\alpha _{2}-\alpha _{3})/3(\alpha _{2}-\alpha _{1})$. For $%
-3a>1$ $\epsilon _{a}=\epsilon _{b}=\epsilon $ and $\bar{\epsilon}_{a}=\bar{%
\epsilon}_{b}=\bar{\epsilon}$. In this case it is easy to check that there
are {\it no} texture zeros because all matrix elements contribute at leading
order to the masses and mixing angles. For $1>-3a>0$, $\epsilon _{a},\;\bar{%
\epsilon}_{a}$ change and are given by $\bar{\epsilon _{a}}=(\frac{<\bar{%
\theta}>}{M_{1}})^{|\alpha _{2}-\alpha _{1}|}$, $\epsilon _{a}=(\frac{<\bar{%
\theta}>}{M_{2}})^{|\alpha _{2}-\alpha _{1}|}$. In this case texture zeros
in the (1,1) and (1,3) positions {\it automatically} appear for small $<\bar{%
\theta}>$. However the (1,2) matrix element is too large (cf Table \ref
{table:2a}). For $a>0$ however $\bar{\epsilon}_{a,b}=(\frac{<\bar{\theta}>}{%
M_{1}})^{|\alpha _{2}-\alpha _{1}|}$, $\epsilon _{a,b}=(\frac{<\bar{\theta}>%
}{M_{2}})^{|\alpha _{2}-\alpha _{1}|}$, the texture zeros in the (1,1) and
(1,3) positions persist, and the (1,2) matrix element can be of the correct
magnitude.

Note that the family symmetry does not make the small elements exactly zero
so it predicts only approximate texture zeros. Indeed, fixing the parameter $%
a=1$ to obtain the measured magnitude of the (1,2) matrix element one finds
that the (1,1) elements occurs at $O(\varepsilon ^{8})$ and $O(\bar{\epsilon}%
^{8})$ for the up and down mass matrices respectively. This is so
small that eq(\ref{eq:vus1}) is valid to a high degree of accuracy if
$\rho_{12}=\rho_{21}$. On the other hand
the (1.3) matrix element is predicted to occur at $O(\epsilon ^{4}),$ $O(%
\bar{\epsilon}^{4})$ for the up and down matrices respectively and, as
discussed above, a term of this order (with a coefficient 2) is sufficient
to correct the prediction for $\left| \frac{V_{ub}}{V_{cb}}\right| $
following from the assumption of an exact texture zero.

However the best fit prefers the (1,3)\ element to occur at $O(\epsilon
^{3}),$ $O(\bar{\epsilon}^{3}),$ i.e. close to the (1,2) matrix elements.
Moreover the measured value of $V_{cb}$ requires the (2,2) and (2,3) matrix
elements of the down quark mass matrices should be of the same magnitude, of
$O(\bar{\epsilon}^{2}).$ This is in contradiction to the predictions of the
Abelian family symmetry, unless one appeals to the unknown coefficients of $%
O(1).$ The most plausible way to get such relations for matrix
elements involving different family members is to invoke a non-Abelian
family symmetry~\cite{tasi}. 

\subsection{Asymmetric case}
\label{sec:u2model}

We now describe a supersymmetric
non-Abelian model based on a U(2) family symmetry
acting on the two lighter families~\cite{Barbieri:1996uv} and on the
unified gauge group SU(5). This model is a
variation~\cite{Barbieri:1997ae} which leads to asymmetric  
textures discussed in Section~\ref{sec:u2} with $Y_{22} \sim 0$.
The lighter families $\psi_a$, $a=1,2$ ($\psi=T,\bar F$, where
$T$ and $\bar F$ are respectively the {\bf 10} and {$\mathbf \bar 5$}
representations of SU(5)) transform as $\psi_a\rightarrow
U_{ab}\psi_b$ under $U\in$U(2), whereas the third family and the Higgs
fields $H_1$, $H_2$ are invariant. Such a symmetry is approximately
realized in nature. In fact, in the U(2) symmetric limit the lighter
fermion families are forced to be massless and in supersymmetric
models their scalar partners are forced to be degenerate. The symmetry
is broken by two SM singlet scalars, an antidoublet $\phi^a$
transforming with $U^{T-1}$ and an antisymmetric tensor $A^{ab}$
transforming with $U^{T-1}\otimes U^{T-1}$ under U(2).  In an
appropriate basis in the flavour space, the corresponding vevs can be
written in the form
\begin{equation}
  \label{vevs}
  \vev{\phi} = \left(
    \begin{array}{c}
      0 \\ V
    \end{array} \right)
\qquad
  \vev{A} = \left(
    \begin{array}{cc}
      0 & v \\
      -v & 0
    \end{array} \right)\; ,
\end{equation}
where $V,v>0$. The correct hierarchy and mixing between the two
lighter families is obtained if $v/V=\ord{|V_{us}|}$. The U(2)
breaking is communicated to the light fermions by an heavy U(2)
anti-doublet $\chi^a$ through a Froggatt-Nielsen mechanism. Under the
gauge group, each $\chi^a$, $a=1,2$, transforms as a full fermion
family, which allows a mixing with the light fermions. The heavy mass
term $M\chi^a \bar\chi_a$ for the fields $\chi^a$ also involves of
course a doublet $\bar\chi_a$ with conjugated tranformations under the
SM and U(2) group. As for the size of the mass term, one simple
possibility is that the scale $M$ is above the SU(5) breaking scale,
$M>M_{\rm GUT}$. A small ratio $V/M$ is then generated if the
U(2) breaking takes place at the SU(5) breaking scale, $V\sim M_{\rm
  GUT}$. Possible SU(5) breaking corrections to the heavy mass $M$
will also be correspondingly smaller. The small ratio $V/M$
determines the small Yukawa couplings accounting for the second
generation masses and mixings, forbidden in the U(2) symmetric limit.
In particular, the correct order of magnitude for the mixing of the
two heavier families is obtained if $V/M\sim\ord{|V_{cb}|}$.

Besides $\chi^a$, $\bar\chi_a$, representing the minimal choice for
the messenger sector, the physics at the GUT scale can involve
additional heavy fields. For example, we have mentioned in the
previous Section the possibility of a mixing in Higgs sector involving
two heavy Higgs fields $H'_1$, $H'_2$, singlets in this case under the
U(2) symmetry as $H_1$, $H_2$. Such a mixing can be used to account
for the hierarchy $m_b\ll m_t$. We therefore include $H'_1$, $H'_2$ in
the model. Since they are allowed to interact both with the light
families and the U(2) breaking sector, the U(2) singlets $H'_1$,
$H'_2$ can also mediate U(2) breaking.  Notice that the scale $M'$ at
which this singlet-mediation takes place is a priori independent of
the scale $M$ associated to the doublet-mediation. For example, if the
mass of the heavy Higgses is set by SU(5) breaking we will have $M'\ll
M$.

At this point one can write the most general renormalizable
superpotential involving the light fermions ($\psi_a$, $\psi_3$), the
Higgs fields ($H_1$, $H_2$), the U(2) breaking fields ($\phi^a$,
$A^{ab}$) and the doublet ($\chi^a$, $\bar\chi_a$) and singlet
($H'_1$, $H'_2$) messengers. Once U(2) is broken, a mixing between the
previously massless fermions and the heavy messengers is generated.
The new light fermions can then be easily identified by diagonalizing
the heavy mass matrix. This leads to the following textures for the up
and down quark mass matrices:
\begin{eqnarray}
  \label{textureD}
  D/h_b &=&
\left( 
  \begin{array}{ccc}
    0 & \epsilon' & 0 \\
    -\epsilon' & 0 & \epsilon \\
    0 & t & 1
  \end{array}
\right) \\
  \label{textureU}
  U/h_t &=&
\left( 
  \begin{array}{ccc}
    0 & c\epsilon\epsilon' & 0 \\
    -c\epsilon\epsilon' & 0 & b\epsilon \\
    0 & a\epsilon & 1
  \end{array}
\right) \; ,
\end{eqnarray}
where $\epsilon = \ord{V/M}$, $\epsilon'=\ord{v/M}$, $t=\ord{V/M'}$
and all other coefficients arise from couplings of order one. One then
obtains the textures~(\ref{u2textureD},\ref{u2textureU}) for the
absolute values of the mass matrices with $\alpha\sim\beta\sim
0$\footnote{Unlike in eqs.~(\ref{u2textureD},\ref{u2textureU}), here
  the $\ord{1}$ coefficients can be complex.}. In particular, the
relation $|D_{12}|=|D_{21}|$ follows from the symmetry properties of
the U(2) representations\footnote{A non negligible correction to
  $|D_{12}|=|D_{21}|$ can arise if $t=\ord{1}$ from the diagonalization
  of the kinetic term.}. 

A few comments are in order. Since we expect
$|D_{23}/D_{33}|=\ord{|V_{cb}|}$, the texture zero in the (2,2)
position requires $t=|D_{32}/D_{33}|\gg|V_{cb}|$ in order to account
for the value of $|D_{32}/D_{33} \cdot D_{23}/D_{33}|\simeq m_s/m_b
=\ord{|V_{cb}|}$. We therefore expect a sizeable $t\gg\epsilon$, which
leads to a non negligible correction to $|V_{ub}/V_{cb}|$. From the
model building point of view, the asymmetry $|D_{32}|\gg |D_{23}|$
corresponding to $t\gg\epsilon$ can be simply accounted for by a
relatively light singlet messenger scale $M'\ll M$ (which is analogous
to the $M_1\ll M_2$ assumption of the Abelian case). In fact, if this
is the case, the leading contribution to $D_{32}$ comes from the
exchange of the U(2) singlets $H'_1$, $H'_2$ at the scale $M'$. As a
consequence, $|D_{32}|$ turns out to be larger than $|D_{23}|$, which
is generated by the exchange of the U(2) doublets $\chi^a$,
$\bar\chi_a$ at the higher scale $M$.  Moreover, $H'_1$, $H'_2$
transform as $\mathbf{5}$ and $\mathbf{\bar 5}$ of SU(5). Therefore,
the singlet exchange at the lower scale $M'$ does not contribute
neither to the (2,3) nor to the (3,2) element in the up quark mass
matrix, so that both $U_{23}$ and $U_{32}$ are of order $\epsilon$.
The larger hierarchy $m_c/m_t \ll m_s/m_b$ follows.  As for the
further suppression of $m_u m_c/m_t^2$ with respect to $m_d
m_s/m_b^2$, that is automatically acheived if $A^{ab}$ is a SU(5)
singlet.  The operator $A^{ab}T_a T_b H$ vanishes in fact in this
case due to the antisymmetry of $A^{ab}$. This is a generic appealing
feature of U(2) models. In order to generate a non-vanishing $U_{12}$
entry, SU(5) breaking effects must be included either in the messenger
masses or through higher dimension operators, thus giving the extra
$\epsilon$ in $U_{12}$.  
}

\section{Conclusions}

The presence of texture zeros in the quark Yukawa matrices can constrain
quite tightly the detailed features of the CKM unitarity triangle. Recent
data has shown that it is no longer viable for $s_{13}$ to be zero and are,
as a result, inconsistent with the most promising texture zero structure.
This result seems quite reliable, following both from the improved bound on
$\Delta m_{Bs}$ as well as the improved value of $|V_{ub}/V_{cb}|$.

Theories invoking family symmetries beyond those of the standard
model can lead to a hierarchal
structure for $U$ and $D$ in which the elements appear in the form $\epsilon
^{k}$, where $\epsilon$ is a small parameter. Motivated by such an
expansion, we have explored a perturbative
approach in which the rotation $s_{13}$ is small but non-zero. One way to do
this is by allowing small entries to replace some of the texture zeros. We
have investigated a common symmetric form
for $U$ and $D$ where, in particular, the (1,3), (3,1) element is no longer
zero. We have also investigated an alternative possibility for generating
$s_{13}$ by allowing an asymmetric form for the (2,3), (3,2) mass matrix
elements.
Both cases lead to a desirable phenomenological result whereby the
perturbation of the ratio $|V_{ub}/V_{cb}|$ from the value
$\sqrt{m_{u}/m_{c}%
}$ is larger than that of $|V_{td}/V_{ts}|$ from $\sqrt{m_{d}/m_{s}}$.

On the theoretical side, we would hope that pinning down the allowed
structures for the Yukawa matrices will provide clues to the nature of an
underlying family symmetry. Our analysis shows that a perturbative expansion
in terms a small parameter is quite successful and therefore supports the
idea that the family symmetry is spontaneously broken at the high energy
scale. This has a concrete realization in the Froggatt Nielsen mechanism
where light and heavy states are mixed via an extension of the `see-saw'
mechanism. If we require the symmetric form of the mass matrices it is
necessary to have non-vanishing (1,3), (3,1) matrix elements. This is an
interesting result because it follows from specific Abelian (and
non-Abelian) family symmetries. Similarly the asymmetric solution can also
be obtained from non-Abelian family symmetries. Improvements in the
measurements of the quark masses and CKM mixing angles and in particular on
$\sin 2\beta$ ( as well as on $\sin 2\alpha$) should help in distinguishing
between these candidate symmetries and possibly lead to a viable theory of
fermion mass generation.

\medskip

\noindent
{\Large \bf Acknowledgments}

L. V-S. would like to thank the ROOT team at CERN for the support on ROOT and
CONACyT (Mexico) for financial support.

\end{document}